\documentclass[12pt]{article}
\usepackage{amsmath}
\usepackage{amssymb}
\numberwithin{equation}{section}

\usepackage[dvips]{graphicx}
\usepackage{epsf}
\usepackage{color}
\usepackage{graphicx}
\usepackage{amsmath}
\usepackage{amssymb}
\usepackage{latexsym}
\usepackage{cite}
\newcommand{\LamGUT}{\Lambda_{\mbox{{\scriptsize G}}}}
\newcommand{\msusy}{m_{\mbox{{\scriptsize SUSY}}}}
\newcommand{\etaCP}{\eta_{\mbox{{\scriptsize CP}}}}
\newcommand{\meff}{m_{\mbox{{\scriptsize eff}}}}
\newcommand{\VTB}{V_{\mbox{{\scriptsize TB}}}}

\begin{document}

\setlength{\baselineskip}{18pt}
\begin{titlepage}
\begin{flushright}
KYUSHU-HET-127
\end{flushright}

\vspace*{1.2cm}
\begin{center}
{\Large\bf Cascade Hierarchy in SUSY {\boldmath $SU(5)$} GUT}
\end{center}
\lineskip .75em
\vskip 1.5cm

\begin{center}
{\large 
Kentaro Kojima$^{a,}$\footnote[1]{E-mail:
\tt kojima@rche.kyushu-u.ac.jp}, 
Hideyuki Sawanaka$^{b,}$\footnote[2]{E-mail:
\tt sawanaka@higgs.phys.kyushu-u.ac.jp}, 
and 
Ryo Takahashi$^{c,}$\footnote[3]{E-mail: 
\tt ryo.takahashi@mpi-hd.mpg.de}
}\\

\vspace{1cm}

$^a${\it Center for Research and Advancement in Higher Education,\\
 Kyushu University, Fukuoka 819-0395, Japan}\\
$^b${\it Department of Physics, Kyushu University, Fukuoka 812-8581, Japan}\\
$^c${\it Max-Planck-Institute f$\ddot{u}$r Kernphysik, Saupfercheckweg 
1, 69117 Heidelberg, Germany}\\

\vspace*{10mm}
{\bf Abstract}\\[5mm]
{\parbox{13cm}{
%%%%%%%%%%%%%%%%%%%%%%%%%%%%%%%%%%%%%%%%%%%%%%%%%%%%%%%%%%%%%%%%
%             ABSTRACT                                         %
%%%%%%%%%%%%%%%%%%%%%%%%%%%%%%%%%%%%%%%%%%%%%%%%%%%%%%%%%%%%%%%%
We study cascade hierarchy in supersymmetric $SU(5)$ grand unified theory. The 
neutrino Dirac mass matrix of the cascade form can lead to the tri-bimaximal 
generation mixing at the leading order in the seesaw mechanism while the down 
quark mass matrix of a hybrid cascade form naturally gives the CKM structure. 
We embed such experimentally favored mass textures into supersymmetric 
$SU(5)$ GUT, which gives a relation between the down quark and charged 
lepton mass matrices. Related phenomenologies, such as lepton flavor 
violating processes and leptogenesis, are also investigated in addition to 
lepton mixing angles.
}}
\end{center}
\end{titlepage}

\tableofcontents

\newpage
%%%%%%%%%%%%%%%%%%%%%%
\section{Introduction} 
%%%%%%%%%%%%%%%%%%%%%%

The current precision measurements of the neutrino oscillation have suggested 
that there are large mixing angles among three generations in the lepton sector 
unlike the quark sector. The experimental data of lepton generation mixing 
angles~\cite{Schwetz:2008er} is well approximated by the tri-bimaximal mixing 
\cite{TBM}, which is given by 
\begin{eqnarray}
  \VTB=
  \left(
    \begin{array}{ccc}
      2/\sqrt{6}  & 1/\sqrt{3} & 0           \\
      -1/\sqrt{6} & 1/\sqrt{3} & -1/\sqrt{2} \\
      -1/\sqrt{6} & 1/\sqrt{3} & 1/\sqrt{2}
    \end{array}
  \right). \label{VTB}
\end{eqnarray}
The characteristic features of this mixing matrix are that the second 
generation of the neutrino mass eigenstate is represented by 
tri-maximal mixture of all flavor eigenstates, 
$\nu_2=\sum_\alpha\nu_\alpha/\sqrt{3}$, and the third generation is 
bi-maximal mixture of $\mu$ and $\tau$ neutrinos, 
$\nu_3=(-\nu_\mu+\nu_\tau)/\sqrt{2}$, in the diagonal basis of the 
charged leptons, respectively. The equation \eqref{VTB} implies the 
following forms of neutrino mass matrix in the flavor basis 
\begin{eqnarray}
  M_\nu=\frac{m_1}{6}
  \left(
    \begin{array}{ccc}
      4  & -2 & -2 \\
      -2 & 1  & 1  \\
      -2 & 1  & 1
    \end{array}
  \right)+\frac{m_2}{3}
  \left(
    \begin{array}{ccc}
      1 & 1 & 1 \\
      1 & 1 & 1 \\
      1 & 1 & 1
    \end{array}
  \right)+\frac{m_3}{2}
  \left(
    \begin{array}{ccc}
      0 & 0  & 0  \\
      0 & 1  & -1 \\
      0 & -1 & 1
    \end{array}
  \right),
\end{eqnarray}
where $m_i$ $(i=1\sim3)$ are the neutrino mass eigenvalues. Such suggestive 
forms of the generation mixing and the neutrino mass matrices give us a 
motivation to look for a flavor structure of the lepton sector. Actually, a 
number of proposals based on a flavor symmetry to unravel it and related 
phenomenologies have been elaborated~\cite{Altarelli:2005yp}. 

Recently, it has been pointed out that the neutrino Dirac mass matrix of a 
cascade form can lead to the tri-bimaximal generation mixing at the leading 
order~\cite{Haba:2008dp}. The mass matrix of the cascade form is given by 
\begin{eqnarray}
  M_{\mbox{{\scriptsize cas}}}=
  \left(
    \begin{array}{ccc}
      \delta & \delta  & \delta \\
      \delta & \lambda & \lambda \\
      \delta & \lambda & 1
    \end{array}
  \right)v,
  \label{cas}
\end{eqnarray}
which is described by small parameters, $|\delta|\ll|\lambda|\ll1$, 
and the dimension-one parameter $v$ denotes an overall mass scale. We call 
this type of hierarchy ``cascade hierarchy'', and the mass matrix with such a 
hierarchy ``cascade matrix''. On the other hand, it is well known that the 
down quark mass matrix of somewhat different hierarchical form, which is given 
by 
\begin{eqnarray}
  M_{\mbox{{\scriptsize hyb}}}=
  \left(
    \begin{array}{ccc}
      \epsilon' & \delta' & \delta' \\
      \delta'   & \lambda' & \lambda' \\
      \delta'   & \lambda' & 1
    \end{array}
  \right)v',
  \label{hyb}
\end{eqnarray}
can explain experimentally observed values of CKM matrix elements, where 
$|\epsilon'|\ll|\delta'|\ll|\lambda'|\ll1$. The $(1,1)$ element of the matrix 
is much smaller than all other elements. However, hierarchical structure is 
similar to the cascade form except for the $(1,1)$ element. In this paper, we 
call this type of hierarchy ``hybrid cascade (H.C.) hierarchy'', and the mass 
matrix with such a hierarchy ``hybrid cascade (H.C.) matrix''. 

The interesting fact that the neutrino Dirac mass matrix of the 
cascade form realize the tri-bimaximal generation mixing and the down 
quark mass matrix of the H.C. form reproduce the CKM structure gives 
us a strong motivation to understand the quark and lepton sectors, 
comprehensively. Towards the comprehensive understanding of the 
quark/lepton sectors, we investigate embedding such hierarchies into a 
supersymmetric grand unified theory (SUSY GUT). In this paper, a case of SUSY 
$SU(5)$ GUT is studied as the simplest example. 

The paper is organized as follows: In section 2, we give more detailed 
explanation about the cascade hierarchies and discuss them with the fermion 
masses and mixing angles. In section 3, we embed the cascade hierarchies into 
a SUSY $SU(5)$ GUT. The texture analyses for the quark/lepton sectors are also 
presented in the section. In section 4, we show the numerical analyses of our 
model, which are the generation mixing angles and related phenomenologies such 
as lepton flavor violating rare decay processes and leptogenesis. In section 5, 
some comments on realizations of this model is presented. Section 6 is devoted 
to the summary. The Appendix gives a discussion about constraints on a 
structure of non-diagonal right-handed neutrino mass matrix. 

%%%%%%%%%%%%%%%%%%%%%%%%%%%%%%%%%%%%%%%%%%%%%%%%%%%%%%%%
\section{Cascade hierarchies and fermion mass matrices}
%%%%%%%%%%%%%%%%%%%%%%%%%%%%%%%%%%%%%%%%%%%%%%%%%%%%%%%%

First we imply the cascade textures to quark and lepton sectors 
independently. The implication of the cascades to the lepton sector 
has been discussed in~\cite{Haba:2008dp}. In the work, both mass 
matrices of neutrino and charged lepton have been taken as the cascade 
form. However, a hybrid type of cascade form would be 
phenomenologically allowed for the charged lepton mass matrix. We 
focus on this point and extend our attention to the quark sector in 
the context of the cascade hierarchies. 

The most famous hierarchical mass (Yukawa) structure is realized by 
the Froggatt-Nielsen (FN) mechanism~\cite{fn}, which is one of 
fascinating approaches to explain the quark mass hierarchy. A lot of 
works based on the FN model have been presented, which are 
typically generating a mass matrix as 
\begin{eqnarray}
  M_{\mbox{{\scriptsize wat}}}=
  \left(
    \begin{array}{ccc}
      \delta^2      & \delta\lambda & \delta \\
      \delta\lambda & \lambda^2     & \lambda \\
      \delta        & \lambda       & 1
    \end{array}
  \right)v.
  \label{wat}
\end{eqnarray}
where $|\delta|\ll|\lambda|\ll1$ and $\mathcal{O}(1)$ coefficients for 
each element have been dropped here. Such a hierarchical form of mass matrix 
can be easily realized by an abelian flavor symmetry. 

On the other hand, the cascade mass matrix shown in \eqref{cas} has 
been recently proposed for the lepton sector~\cite{Haba:2008dp}. 
Such a kind of mass matrix can realize the tri-bimaximal generation 
mixing at the leading order in the framework of the seesaw mechanism 
\cite{seesaw} when the hierarchy of Dirac mass matrices for leptons is 
described as the cascade form. In the work, mass matrices of right-handed 
neutrino and charged lepton have been taken as diagonal and cascade form, 
respectively, and some collections from the charged lepton to the tri-bimaximal 
mixing have also been discussed. However, if a right-handed Majorana mass 
matrix gives only small mixing collections, it is experimentally allowed, even 
if it does not have a diagonal form. Therefore, we can conclude for the forms 
of neutrino mass matrices that the tri-bimaximal generation mixing can be 
realized at the leading order if the Dirac mass matrices of the cascade form 
and a Majorana mass matrix leading small mixing collections are taken. This 
means that the cascade or H.C. form of Majorana mass matrix are also allowed. 
The form of charged lepton mass matrix does not have to be the cascade form; 
if the matrix also induces only small collections to the mixing angles such as 
the H.C. form, it is allowed. 

The mass matrix shown in~\eqref{wat} has a more rapid stream of 
hierarchy flow than the cascade one given in~\eqref{cas}, and is 
called the ``waterfall'' mass matrix. We note that these two types of 
matrices have the same orders of generation mixing angles, while they 
induce different mass eigenvalues as shown in Tab.~\ref{tab1}. 
\begin{table}[t]
\begin{center}
\begin{tabular}{|c|c|c|}
  \hline
  & cascade & waterfall \\
  \hline
  \hline
  mass eigenvalues & $m_1:m_2:m_3\sim\delta:\lambda:1$ 
  & $m_1:m_2:m_3\sim\delta^2:\lambda^2:1$ \\
  \hline
  mixing angles 
  & $\theta_{12}\sim\delta/\lambda$, $\theta_{23}\sim\lambda$, $\theta_{13}\sim\delta$ 
  & $\theta_{12}\sim\delta/\lambda$, $\theta_{23}\sim\lambda$, $\theta_{13}\sim\delta$ \\
  & ($\theta_{ij}\sim m_i/m_j$) & ($\theta_{ij}\sim\sqrt{m_i/m_j}$) \\
  \hline
\end{tabular}
\end{center}
\caption{Mass eigenvalues and generation mixing angles induced from 
the cascade and waterfall mass matrices.}
\label{tab1}
\end{table}
It is seen that relation among the mass eigenvalues and mixing angles 
are roughly estimated as $\theta_{ij}\sim m_i/m_j$ for the cascade 
matrix and $\theta_{ij}\simeq\sqrt{m_i/m_j}$ for the waterfall. 

Note that the experimentally observed values of masses and mixing 
angles in the quark sector are well approximated by 
\begin{eqnarray}
  \theta_{12}^q\sim\sqrt{\frac{m_{d_1}}{m_{d_2}}},~~
  \theta_{23}^q\sim\frac{m_{d_2}}{m_{d_3}},~~
  \theta_{13}^q\sim\theta_{12}^q\theta_{23}^q
  \sim{\frac{\sqrt{m_{d_1}m_{d_2}}}{m_{d_3}}},
\end{eqnarray}
where $m_{d_i}$ are the down-type quark masses. We find that 
$\theta_{12}^q$ and $\theta_{23}^q$ can be realized in the waterfall 
and cascade matrix, respectively. That motivates us to combine these 
matrices such that all mixing angle are appropriately obtained. It can 
be achieved by a hybrid form of mass matrix given in \eqref{hyb}. 
The induced mass eigenvalues and mixing angles from the H.C. 
matrix are shown in Tab.~\ref{tab2}. 
\begin{table}[t]
\begin{center}
\begin{tabular}{|c|c|}
  \hline
  & hybrid cascade  \\
  \hline
  \hline
  mass eigenvalues & $m_1:m_2:m_3\sim\delta^2/\lambda:\lambda:1$ \\
  \hline
  mixing angles & $\theta_{12}\sim\delta/\lambda$, $\theta_{23}\sim\lambda$, $\theta_{13}\sim\delta$ \\
  & ($\theta_{12}\sim\sqrt{m_1/m_2}$, $\theta_{23}\sim m_2/m_3$, $\theta_{13}\sim\sqrt{m_1m_2}/m_3$) \\
  \hline
\end{tabular}
\end{center}
\caption{Mass eigenvalues and generation mixing angles induced from the hybrid 
cascade mass matrix, where we replace $\delta'$ and $\lambda'$ in \eqref{hyb} 
with $\delta$ and $\lambda$, respectively.}
\label{tab2}
\end{table}
It is remarked that while the structure of the 2nd--3rd sector is same 
for cascade and H.C mass matrices, the magnitude of hierarchy 
between mass eigenvalues of the 1st and 2nd generation in the H.C. 
matrix is larger than that of the cascade matrix. Since the mass hierarchy 
of the up-type sector is much larger than that of the down-type 
sector, the CKM matrix is almost determined by the structure of mass 
matrix for the down-type quarks. Therefore, the mass hierarchy of the 
down-type quark mass matrix should be taken as the H.C. form. Both 
cascade and H.C. forms can be taken as the mass matrix for the up-type 
sector because the contributions from the up-type sector to the CKM 
mixing angles are small compared with that from the down-type 
sector. Moreover, an arbitrary form of up quark mass matrix is allowed 
as long as collections from the matrix is enough small. It is seen 
that if the mass matrix for the up sector is described as the cascade form, 
a larger hierarchy between $\delta$ and $\lambda$ than that of the 
H.C. case is needed. These discussions for the cascading fermion mass 
matrices are summarized in Tab.~\ref{tab3}. 
\begin{table}[t]
\begin{center}
\begin{tabular}{|c||c|}
  \hline
  & mass textures \\
  \hline
  \hline
  up-type quark & $M_u: \mbox{cascade\ or H.C. or small mixing}$ \\
  \hline
  down-type quark & $M_d: \mbox{H.C.}$ \\
  \hline
  \hline
  neutrino Dirac & $M_{\nu D}: \mbox{cascade}$ \\
  \hline 
  charged lepton & $M_e: \mbox{cascade\ or H.C. or small mixing}$ \\
  \hline
  \hline
  right-handed Majorana & $M_R: \mbox{cascade\ or H.C. or small mixing}$\\
  \hline 
\end{tabular}
\end{center}
\caption{Experimentally allowed mass textures for the fermion mass matrices 
based on cascading hierarchies.}
\label{tab3}
\end{table}

%%%%%%%%%%%%%%%%%%%%%%%%%%%%%%%%%%%%%%%%%%%%
\section{Cascade hierarchies in {\boldmath $SU(5)$} GUT}
%%%%%%%%%%%%%%%%%%%%%%%%%%%%%%%%%%%%%%%%%%%%

In this section, we consider embedding the cascade textures into 
SUSY $SU(5)$ GUT and realizations in the theory. One parameter fit for the 
cascading hierarchies is also discussed. 

%%%%%%%%%%%%%%%%%%%%%%%%%%%%%%%%%%%%%%%%%%
\subsection{Fermion masses in $SU(5)$ GUT}
%%%%%%%%%%%%%%%%%%%%%%%%%%%%%%%%%%%%%%%%%%

We give a brief review of the fermion masses in $SU(5)$ GUT before 
considering the cascade textures in the theory. 

In the $SU(5)$ GUT model, the SM fermions belong to the following 
representations, 
\begin{eqnarray}
  \bar{\psi}_i=
  \left(
    \begin{array}{c}
      d_1^c \\
      d_2^c \\
      d_3^c \\
      e^-   \\
      \nu  
    \end{array}
  \right)_L,~~~
  \psi_{ij}=
  \left(
    \begin{array}{ccccc}
      0 & u_3^c & -u_2^c & u_1 & d_1 \\
      -u_3^c & 0 & u_1^c & u_2 & d_2 \\
      u_2^c & -u_1^c & 0 & u_3 & d_3 \\
      -u_1 & -u_2 & -u_3 & 0 & e^+ \\
      -d_1 & -d_2 & -d_3 & -e^+ & 0
    \end{array}
  \right)_L,~~~
  \psi_0=\nu_R^c,
\end{eqnarray}
where the indices $i$ and $j$ ($i,j=1\sim3$) correspond to the color 
ones and $i,j=4,5$ are the weak isospin $I_3=+1/2$ and $-1/2$, 
respectively. The matter fields $\bar{\psi}_i$ and $\psi_{ij}$ are 
transformed as $\bar{5}$ and $10$ representations of $SU(5)$, respectively. 
The right-handed neutrino can be introduced as the singlet under the gauge 
group. The Higgs fields are assigned to $H_{45}$, $H_5$ and $\bar{H}_5$, 
which are transformed as the $\bar{45}$, $5$ and $\bar{5}$ representations. 
It is seen that a relation between mass matrices of the charged lepton and 
down-type quark, $M_e\simeq M_d^T$, is induced as a characteristic prediction 
of the $SU(5)$ GUT since the down-type quarks and charged leptons belong 
to the same representation. We will discuss the cascade hierarchical 
mass matrices in the $SU(5)$ GUT while considering such a relation in 
the following section. 

%%%%%%%%%%%%%%%%%%%%%%%%%%%%%%%%%%%%%%%%%%%%%%%%%%%%%%%%
\subsection{Possible types of the cascade hierarchies}
%%%%%%%%%%%%%%%%%%%%%%%%%%%%%%%%%%%%%%%%%%%%%%%%%%%%%%%%

Let us argue embedding the (hybrid) cascade hierarchical mass matrices 
into $SU(5)$ GUT. The $SU(5)$ GUT predicts a relation between mass 
matrices for the down-type quark and charged lepton, $M_e\simeq M_d^T$, 
due to the unification of matter contents. As discussed above, 
since only the mass matrix of the H.C. form are allowed for $M_d$, 
the mass matrix for the charged lepton should also have the 
H.C. form. On the other hand, some hierarchical structure of the mass 
matrices for the up-type quark $M_u$ and the right-handed neutrino 
$M_R$ are allowed as long as induced mixing angles from these matrices 
can be treated as collections for the CKM and PMNS matrices, 
respectively. It is seen that the (3,3) element of Yukawa matrix for 
$M_u$ should be of order one, $(Y_u)_{33}\sim\mathcal{O}(1)$, while 
$(Y_\ast)_{33}\ll1$ are generally allowed for $Y_\ast$ $(\ast=d,\nu,e)$, which 
correspond to the Yukawa matrices for the down-type quark, neutrino, and 
charged lepton, respectively. Therefore, we parametrize the mass matrices of 
the cascade or H.C. form for the fermions towards the realization of 
embedding the cascading texture into $SU(5)$ GUT as 
\begin{eqnarray}
  M_u &\simeq& 
  \left(
    \begin{array}{ccc}
      \epsilon_u & \delta_u  & \delta_u \\
      \delta_u   & \lambda_u & \lambda_u \\
      \delta_u   & \lambda_u & 1
    \end{array}
  \right)v_u,\hspace{3mm}\mbox{ with }
  \left\{
    \begin{array}{lll}
      |\epsilon_u|=|\delta_u|\ll|\lambda_u|\ll1   & : & \mbox{cascade} \\
      |\epsilon_u|\ll|\delta_u|\ll|\lambda_u|\ll1 & : & \mbox{H.C.}
    \end{array}
  \right.,\label{up}\\
  M_d &\simeq& 
  \left(
    \begin{array}{ccc}
      \epsilon_d & \delta_d  & \delta_d \\
      \delta_d   & \lambda_d & \lambda_d \\
      \delta_d   & \lambda_d & 1
    \end{array}
  \right)\xi_dv_d,\hspace{0.5mm}\mbox{ with } |\epsilon_d|\ll|\delta_d|\ll|\lambda_d|\ll1:\mbox{H.C.},\label{down} \\
  M_{\nu D} &\simeq&
  \left(
    \begin{array}{ccc}
      \delta_\nu & \delta_\nu  & \delta_\nu \\
      \delta_\nu & \lambda_\nu & \lambda_\nu \\
      \delta_\nu & \lambda_\nu & 1
    \end{array}
  \right)\xi_\nu v_u,\mbox{ with } |\delta_\nu|\ll|\lambda_\nu|\ll1:\mbox{cascade}, \label{neu}\\
  M_e &\simeq& 
  \left(
    \begin{array}{ccc}
      \epsilon_e & \delta_e  & \delta_e  \\
      \delta_e   & \lambda_e & \lambda_e \\
      \delta_e   & \lambda_e & 1
    \end{array}
  \right)\xi_ev_d,\hspace{1mm}\mbox{ with } |\epsilon_e|\ll|\delta_e|\ll|\lambda_e|\ll1:\mbox{H.C.},\label{charged}\\
  M_R &\simeq& 
  \left(
    \begin{array}{ccc}
      \epsilon_R & \delta_R  & \delta_R \\
      \delta_R   & \lambda_R & \lambda_R \\
      \delta_R   & \lambda_R & 1
    \end{array}
  \right)M,\hspace{1.5mm}\mbox{ with } 
  \left\{
    \begin{array}{lll}
      |\epsilon_R|=|\delta_R|\ll|\lambda_R|\ll1   & : & \mbox{cascade} \\
      |\epsilon_R|\ll|\delta_R|\ll|\lambda_R|\ll1 & : & \mbox{H.C.}
    \end{array}
  \right.,
\end{eqnarray}
where overall factor $\xi_*$ $(*=d,\nu,e)$ is at most ${\cal O}(1)$, and 
for the matrix elements $\mathcal{O}(1)$ coefficients have been 
dropped. Vacuum expectation values of up- and down-type Higgs fields in the 
supersymmetric scenario are shown by $v_u$ and $v_d$. The characteristic 
relation from $SU(5)$ GUT, $M_e\simeq M_d^T$, is discussed in the next 
subsection. 
 
%%%%%%%%%%%%%%%%%%%%%%%%%%%%%%%%%%%%%%%%%%%%%%%%%%%%%%%%
\subsection{One-parameter fit of cascade mass matrices}
\label{sec-op}
%%%%%%%%%%%%%%%%%%%%%%%%%%%%%%%%%%%%%%%%%%%%%%%%%%%%%%%%

The cascade parameters in the quark and charged lepton sectors, 
$\epsilon_y$, $\delta_y$, and $\lambda_y$ $(y=u,d,e)$, can be estimated 
by experimental values. As shown in~\eqref{up},~\eqref{down} 
and~\eqref{charged}, the mass matrix of the up-type quark can be 
described by either the cascade or H.C. types of hierarchies while only 
the mass matrix of the H.C. form is allowed for the mass matrices of 
the down-type and charged lepton. Typical magnitudes of cascade 
parameters at a low-energy regime and the GUT scale are shown in 
Tab.~\ref{tab5}. 
\begin{table}[t]
\begin{center}
\begin{tabular}{|c||c|c|}
\hline
$M_u: \mbox{cascade}$ & Low-energy scale & $\mathcal{O}(10^{16})\mbox{ GeV}$ \\
\hline
\hline
$\lambda_u\simeq m_c/m_t$ & $7.1\times10^{-3}$ & $2.3\times10^{-3}$ \\
\hline
$\delta_u\simeq m_u/m_t$ & $1.1\times10^{-5}$ & $6.0\times10^{-6}$ \\
\hline
$\theta_{u,12}\simeq\delta_u/\lambda_u\simeq m_u/m_c$ & $1.6\times10^{-3}$ & $2.6\times10^{-3}$ \\
\hline
$\theta_{u,23}\simeq\lambda_u\simeq m_c/m_t$ & $7.1\times10^{-3}$ & $2.3\times10^{-3}$ \\
\hline
$\theta_{u,13}\simeq\delta_u\simeq m_u/m_t$ & $1.1\times10^{-5}$ & $6.0\times10^{-6}$ \\
\hline
\hline
$M_u: \mbox{H.C.}$ & & \\
\hline
\hline
$\lambda_u\simeq m_c/m_t$ & $7.1\times10^{-3}$ & $2.3\times10^{-3}$ \\
\hline
$\delta_u\simeq\sqrt{m_um_c}/m_t$ & $2.8\times10^{-4}$ & $1.2\times10^{-4}$ \\
\hline
$\theta_{u,12}\simeq\delta_u/\lambda_u\simeq\sqrt{m_u/m_c}$ & $4.0\times10^{-2}$ & $5.1\times10^{-2}$ \\
\hline
$\theta_{u,23}\simeq\lambda_u\simeq m_c/m_t$ & $7.1\times10^{-3}$ & $2.3\times10^{-3}$ \\
\hline
$\theta_{u,13}\simeq\delta_u\simeq\sqrt{m_um_c}/m_t$ & $2.8\times10^{-4}$ & $1.2\times10^{-4}$ \\
\hline
\hline
$M_d: \mbox{H.C.}$ & & \\
\hline
\hline
$\lambda_d\simeq m_s/m_b$ & $2.4\times10^{-2}$ & $1.8\times10^{-2}$ \\
\hline
$\delta_d\simeq\sqrt{m_dm_s}/m_b$ & $5.4\times10^{-3}$ & $4.1\times10^{-3}$ \\
\hline
$\theta_{d,12}\simeq\delta_d/\lambda_d\simeq\sqrt{m_d/m_s}$ & $0.23$ & $0.23$ \\
\hline
$\theta_{d,23}\simeq\lambda_d\simeq m_s/m_b$ & $2.4\times10^{-2}$ & $1.8\times10^{-2}$ \\
\hline
$\theta_{d,13}\simeq\delta_d\simeq\sqrt{m_dm_s}/m_b$ & $5.4\times10^{-3}$ & $4.1\times10^{-3}$ \\
\hline
\hline
$M_e: \mbox{H.C.}$ & & \\
\hline
\hline
$\lambda_e\simeq m_\mu/m_\tau$ & $5.9\times10^{-2}$ & $5.4\times10^{-2}$ \\
\hline
$\delta_e\simeq\sqrt{m_em_\mu}/m_\tau$ & $4.0\times10^{-3}$ & $3.7\times10^{-3}$ \\
\hline
$\theta_{e,12}\simeq\delta_e/\lambda_e\simeq\sqrt{m_e/m_\mu}$ & $6.8\times10^{-2}$ & $6.9\times10^{-2}$ \\
\hline
$\theta_{e,23}\simeq\lambda_e\simeq m_\mu/m_\tau$ & $5.9\times10^{-2}$ & $5.4\times10^{-2}$ \\
\hline
$\theta_{e,13}\simeq\delta_e\simeq\sqrt{m_em_\mu}/m_\tau$ & $4.0\times10^{-3}$ & $3.7\times10^{-3}$ \\
\hline
\end{tabular}
\end{center}
\caption{Typical magnitude of cascade parameters at a low-energy regime and 
the GUT scale.}
\label{tab5}
\end{table}
The $u$, $d$, and $s$ quark masses are estimations of current-quark 
mass in a $\overline{\mbox{MS}}$ scheme at a scale $\mu=2$ GeV 
\cite{pdg}. The $c$ and $b$ quark masses are the running masses in the 
scheme. The top quark mass is determined by the direct observation of 
top events. In the supersymmetric scenario, threshold effects arise 
from decoupling of the supersymmetric partner of SM particles could 
play an important role to determine the GUT scale fermion masses 
(Yukawa couplings)~\cite{deltab,gutph}. In particular, a wide region 
of $b$ quark mass has been considered because it strongly depends on 
the low-energy SUSY threshold effects. In the analysis focusing on the 
cascade texture, we refer to a typical GUT scale mass parameters 
listed in~\cite{Ross:2007az} where the Georgi-Jarlskog (GJ) 
factor~\cite{Georgi:1979df} successfully explain down-type quark and 
charged lepton mass spectrum, as seen below. Here, we give some 
comments from the Tab.~\ref{tab5}: 
\begin{itemize}
\item When $M_u$ takes the cascade form, effects on the CKM mixing from 
the up-type quark sector are little. The CKM mixing is almost determined 
by the structure of down-type quark mass matrix. 

\item There are collections of $\mathcal{O}(10\%)$ from the up sector 
to the CKM mixing when $M_u$ takes the H.C. form. 

\item It is known that the mass ratio between the down-type quarks and charged 
leptons for each generation can be written as 
\begin{eqnarray}
  \left(\frac{m_\tau}{m_b},\frac{m_\mu}{m_s},\frac{m_e}{m_d}\right)
  \sim\left(1,3,\frac{1}{3}\right)
\end{eqnarray}
Therefore, $\theta_{e,23}$ is larger than $\theta_{d,23}$ while 
$\theta_{e,12}$ is smaller than $\theta_{d,12}$. Throughout this 
paper, we introduce the GJ factor~\cite{Georgi:1979df} in the charged 
lepton mass matrix. If the (2,2) element of the Yukawa matrices is 
generated by the operator $\bar{5}\cdot10\cdot H_{45}$ with the 
standard model (SM) Higgs fields contained in the 45-dimensional 
representation $H_{45}$, where $\bar{5}$ and $10$ stand for matters 
described by the 5- and 10-dimensional representations, 
respectively. The operator leads to the well-known relation $m_\mu/m_s=-3$, 
which is favored by the experimental data. It can be understood 
from the fact that the 45-dimensional representation is traceless and 
the factor of $-3$ for the charged leptons, and thus, it has to 
compensate the color factor of $3$ for the quarks. Hereafter we 
express the mass matrices of the down-type quark and charged lepton 
replaced with~\eqref{charged} and~\eqref{down} as 
\begin{eqnarray}
  M_e\simeq 
  \left(
    \begin{array}{ccc}
      \epsilon_d & \delta_d    & \delta_d \\
      \delta_d   & -3\lambda_d & \lambda_d \\
      \delta_d   & \lambda_d   & 1
    \end{array}
  \right)\xi_dv_d,~~~
  M_d\simeq 
  \left(
    \begin{array}{ccc}
      \epsilon_d & \delta_d  & \delta_d \\
      \delta_d   & \lambda_d & \lambda_d \\
      \delta_d   & \lambda_d & 1
    \end{array}
  \right)\xi_dv_d,\label{GJ}
\end{eqnarray}
by using the same cascade parameters $\epsilon_d$, $\delta_d$, and 
$\lambda_d$. Furthermore, the fact $m_\tau/m_b\sim1$ leads to 
$\xi_e\sim\xi_d$. 
\end{itemize}

There is no relation between the hierarchies of up- and down-type 
quarks (charged lepton) in the context of the $SU(5)$ GUT. However it 
is natural to expect that such hierarchies are originated from a 
symmetry and/or some dynamics in a high energy regime rather than 
solely determined by the magnitudes of Yukawa coupling constants. Here 
we express the cascade parameters shown in Tab.~\ref{tab5} by an unit 
of the Cabibbo angle, $\sin\theta_c\simeq\lambda\simeq0.227$ in 
Tab.~\ref{tab6}. 
\begin{table}
\begin{center}
\begin{tabular}{|c||c|c|}
\hline
$M_u:$ cascade\ & Low-energy scale & $\mathcal{O}(10^{16})$ GeV \\
\hline
$\lambda_u$ & $0.61\times\lambda^4$ & $0.87\times\lambda^4$  \\
$\delta_u$  & $0.35\times\lambda^8$ & $0.85\times\lambda^8$ \\
\hline
\hline
$M_u:$ H.C.  & & \\
\hline
$\lambda_u$ & $0.61\times\lambda^4$ & $0.87\times\lambda^4$  \\
$\delta_u$  & $0.46\times\lambda^6$ & $0.86\times\lambda^6$ \\
\hline
\hline
$M_d:$ H.C.  & & \\
\hline
$\lambda_d$ & $0.47\times\lambda^2$ & $0.35\times\lambda^2$  \\
$\delta_d$  & $0.46\times\lambda^3$ & $0.35\times\lambda^3$ \\
\hline
\end{tabular}
\end{center}
\caption{Typical magnitudes of cascade parameters in an unit of the Cabibbo 
angle, $\sin\theta_c\simeq\lambda\simeq0.227$.}
\label{tab6}
\end{table}
Here we have taken the $\mathcal{O}(1)$ coefficient of (3,3) element in 
each matrix as one. If the coefficient is taken as 
$a\sim\mathcal{O}(1)$, other coefficients in the same matrix are 
multiplied by a factor $a$. 

We note that the since $\xi_d$ determines a ratio between (3,3) element of 
Yukawa matrices for up- and down-type quarks, it is correlated with the 
$\tan\beta$, which is the ratio between vacuum expectation values of up- and 
down-type Higgs fields in the supersymmetric scenario. 
For a suppressed value of $\xi_d$ leads to a small $\tan\beta$ as follows, 
\begin{eqnarray}
  \tan\beta\simeq
  \left\{
    \begin{array}{llll}
      v_u/v_d\sim\mathcal{O}(50) & \mbox{ for } & \xi_d\sim\lambda^0 & \mbox{ [large]}\\
      \lambda v_u/v_d\sim\mathcal{O}(10) & \mbox{ for } & \xi_d\sim\lambda^1 & \mbox{ [moderate]}\\
      \lambda^2v_u/v_d\sim\mathcal{O}(1) & \mbox{ for } & \xi_d\sim\lambda^2 & \mbox{ [small]} 
    \end{array}
  \right..
\end{eqnarray}

It is also remarked that the magnitude of hierarchy for the 1st 
generation in the up-type quark sector should be large for the 
cascade form of $M_u$, $\delta_u\sim\mathcal{O}(\lambda^8)$, compared to the 
down-type quark sector, $\delta_d\sim\mathcal{O}(\lambda^3)$. 
On the other hand, H.C. form of $M_u$ has milder hierarchy than cascade one, 
and similar stream of hierarchy flow to $M_d$. 
From a viewpoint of some flavor model, like FN model, the different 
hierarchies could be accompanied by properties of $10$ and $\bar{5}$ matter 
fields, since mass matrices $M_u$ and $M_d$ are accompanied by $10\cdot10$ and 
$10\cdot\bar{5}$ matter complings, respectively. 
To differ the hierarchy of $M_u$ from that of $M_d$ drastically, 
a specific (unnatural) mechanism for the difference is expected. 
The realization may be possible but we focus on the case of 
the H.C. mass matrix for $M_u$ in the following discussions. 

Finally, we can take cascading textures at GUT scale as 
\begin{eqnarray}
  M_u &\simeq& 
  \left(
    \begin{array}{ccc}
      \lambda^{k_u+6} & \lambda^6 & \lambda^6 \\
      \lambda^6       & \lambda^4 & \lambda^4 \\
      \lambda^6       & \lambda^4 & 1
    \end{array}
  \right)v_u, \label{cas-mu}\\
  M_d &\simeq& 
  \left\{
    \begin{array}{ll}
      \left(
        \begin{array}{ccc}
          \lambda^{k_d+3} & \lambda^3 & \lambda^3 \\
          \lambda^3       & \lambda^2 & \lambda^2 \\
          \lambda^3       & \lambda^2 & 1
        \end{array}
      \right)v_d & \mbox{[large $\tan\beta$]} \\
      \left(
        \begin{array}{ccc}
          \lambda^{k_d+4} & \lambda^4 & \lambda^4 \\
          \lambda^4       & \lambda^3 & \lambda^3 \\
          \lambda^4       & \lambda^3 & \lambda
        \end{array}
      \right)v_d & \mbox{[moderate $\tan\beta$]} \\
      \left(
        \begin{array}{ccc}
          \lambda^{k_d+5} & \lambda^5 & \lambda^5 \\
          \lambda^5       & \lambda^4 & \lambda^4 \\
          \lambda^5       & \lambda^4 & \lambda^2
        \end{array}
      \right)v_d & \mbox{[small $\tan\beta$]} \\
    \end{array}
  \right.,\label{cas-md}
\end{eqnarray}
where $k_u\geq2$ and $k_d\geq1$ are needed to obtain suitable mass 
eigenvalues after diagonalizing these matrices. It should be 
remembered that $M_e\simeq M_d^T$ but the additional GJ factor $-3$ is 
multiplied to the (2,2) element of $M_e$ as discussed in~\eqref{GJ}. 

%%%%%%%%%%%%%%%%%%%%%%%%%%%%%%%%%%%
\subsection{Neutrino sector}\label{sec-Neutrino-sector}
%%%%%%%%%%%%%%%%%%%%%%%%%%%%%%%%%%%

Next, we discuss the neutrino mass matrix with cascading form. The 
neutrino Dirac mass matrix must be taken as the cascade form to lead 
to a nearly tri-bimaximal generation mixing. In order to realize the 
tri-bimaximal pattern, mixing angles among the right-handed neutrinos 
should be small. The Majorana mass matrix of the right-handed neutrinos 
$M_R$ has been taken to be diagonal in~\cite{Haba:2008dp}. In the 
case, cascade parameters are constrained as 
\begin{eqnarray}
  \left|\frac{\delta_\nu}{\lambda_\nu}\right|^2
  \ll\frac{\Delta m_{21}^2}{|\Delta m_{31}^2|}
  \simeq3.19\times10^{-2}<\lambda^2,
  \label{constraint}
\end{eqnarray}
to conserve the tri-bimaximal mixing at the leading order. Here 
$\Delta m_{21}^2\equiv|m_2|^2-|m_1|^2$ and 
$\Delta¡¡m_{31}^2\equiv|m_3|^2-|m_1|^2$ are the mass squared 
difference of light neutrinos and the current experimental data at the 
$3\sigma$ level~\cite{Schwetz:2008er} are 
\begin{eqnarray}
  \Delta m_{21}^2 &=& (7.695\pm0.645)\times10^{-5}\mbox{ eV}^2, \label{sol}\\
  |\Delta m_{31}^2| &=& 2.40_{-0.11}^{+0.12}\times10^{-3}\mbox{ eV}^2. \label{atm}
\end{eqnarray}
The equation~ (\ref{constraint}) constrains the magnitude of the 
cascade hierarchies for neutrino Dirac mass matrix to be 
$|\delta_\nu/\lambda_\nu|<\lambda$. If we assume that the hierarchy in 
the Dirac mass matrix is originated from the $\lambda$,\footnote{As 
commented below, if the origin of hierarchy in the neutrino sector 
is assumed to be same as one in the quark sector, the cascade 
parameter in the neutrino sector could also be described by a 
Cabibbo unit. Such a situation is considered to be natural in a realization of 
the cascade and H.C. textures due to the FN mechanism with an 
abelian flavor symmetry such as $U(1)$.} we can 
reparametrize the cascade matrix~(\ref{neu}) as 
\begin{eqnarray}
  M_{\nu D} &\simeq&
  \left(
    \begin{array}{ccc}
      \lambda^{d_1} & \lambda^{d_1} & \lambda^{d_1} \\
      \lambda^{d_1} & \lambda^{d_2} & -\lambda^{d_2} \\
      \lambda^{d_1} & -\lambda^{d_2} & 1
    \end{array}
  \right)\lambda^dv_u\hspace{3mm}\mbox{ with }\hspace{3mm}d_1>d_2\geq1
  \hspace{3mm}\mbox{ and }\hspace{3mm}d\geq0,\label{Dirac}
\end{eqnarray}
where $d_1$ and $d_2$ determine the magnitudes of hierarchy, and an 
opposite sign between (2,2) and (2,3) elements is experimentally 
required as commented in~\cite{Haba:2008dp}. Here we constrain $d$, 
$d_1$, and $d_2$ to be integer. The considerations about constraints 
on the magnitude of cascade hierarchy of the neutrino Dirac mass 
matrix in this parametrization will be discussed in the following 
subsections. 

%%%%%%%%%%%%%%%%%%%%%%%%%%%%%%%%
\subsubsection{Diagonal $M_R$ case}
%%%%%%%%%%%%%%%%%%%%%%%%%%%%%%%%

First, let us consider a case of diagonal Majorana mass matrix of the 
right-handed neutrinos, $M_R$.\footnote{This case has been proposed in 
\cite{Haba:2008dp}. Here we give a brief review of the work and constraints on 
cascade parameters in our notation.} We take a diagonal 
form of Majorana mass matrix as, 
\begin{eqnarray}
  M_R\simeq
  \left(
    \begin{array}{ccc}
      \lambda^{x_1} & 0 & 0 \\
      0 & \lambda^{x_2} & 0 \\
      0 & 0 & 1
    \end{array}
  \right)M \hspace{3mm}\mbox{ with }\hspace{3mm}x_1\geq x_2\geq0,
\end{eqnarray}
where $M$ is a mass scale of the heaviest right-handed neutrino. After the 
seesaw mechanism, one obtains the Majorana mass matrix of light neutrinos in 
low-energy theory, 
\begin{eqnarray}
  M_\nu &\simeq& \left[
    \left(
      \begin{array}{ccc}
        \lambda^{2d_1}     & -\lambda^{d_1+d_2}  & \lambda^{d_1}  \\
        -\lambda^{d_1+d_2} & \lambda^{2d_2}      & -\lambda^{d_2} \\
        \lambda^{d_1}      & -\lambda^{d_2}      & 1
      \end{array}
    \right)+\lambda^{2d_1-x_1}
    \left(
      \begin{array}{ccc}
        1 & 1 & 1 \\
        1 & 1 & 1 \\
        1 & 1 & 1
      \end{array}
    \right)\right.\nonumber\\
  &&\phantom{\Bigg[}\left.+\lambda^{2d_2-x_2}
    \left(
      \begin{array}{ccc}
        \lambda^{2(d_1-d_2)} & \lambda^{d_1-d_2} & -\lambda^{d_1-d_2} \\
        \lambda^{d_1-d_2}    & 1                 & -1                 \\
        \lambda^{d_1-d_2}    & -1                & 1
      \end{array}
    \right)\right]\frac{\lambda^{2d}v_u^2}{M}.\label{neu-mass}
\end{eqnarray}
Operating the $V_{\mbox{{\scriptsize TB}}}$ to $M_\nu$ as 
$V_{\mbox{{\scriptsize TB}}}^TM_\nu V_{\mbox{{\scriptsize TB}}}$, the resultant 
neutrino mass matrix becomes 
\begin{eqnarray}
  \mathcal{M}&\equiv&V_{\mbox{{\scriptsize TB}}}^TM_\nu 
  V_{\mbox{{\scriptsize TB}}}\nonumber\\
  &\simeq&\left[\frac{1}{6}\left(
      \begin{array}{ccc}
        c_1^2           & -\sqrt{2}c_1c_2 & -\sqrt{3}c_1c_+ \\
        -\sqrt{2}c_1c_2 & 2c_2^2          & \sqrt{6}c_2c_+  \\
        -\sqrt{3}c_1c_+ & \sqrt{6}c_2c_+  & 3c_+^2
      \end{array}\right)+3\lambda^{2d_1-x_1}\left(
      \begin{array}{ccc}
        0 & 0 & 0 \\
        0 & 1 & 0 \\
        0 & 0 & 0
      \end{array}\right)\right.\nonumber\\
  &&\phantom{\Bigg[}\left.+\frac{\lambda^{2d_2-x_2}}{3}\left(
      \begin{array}{ccc}
        2\lambda^{2(d_1-d_2)} & \sqrt{2}\lambda^{2(d_1-d_2)} & -2\sqrt{3}\lambda^{d_1-d_2} \\
        \sqrt{2}\lambda^{2(d_1-d_2)} & \lambda^{2(d_1-d_2)} & -\sqrt{6}\lambda^{d_1-d_2} \\
        -2\sqrt{3}\lambda^{d_1-d_2} & -\sqrt{6}\lambda^{d_1-d_2} & 6
      \end{array}\right)\right]\frac{\lambda^{2d}v_u^2}{M},
  \label{neu-mass-diag}
\end{eqnarray}
where 
\begin{eqnarray}
  c_1  &\equiv&1-\lambda^{d_2}-2\lambda^{d_1}, \\
  c_2  &\equiv&1-\lambda^{d_2}+\lambda^{d_1},  \\
  c_3  &\equiv&1-\lambda^{d_2}+4\lambda^{d_1}, \\
  c_+&\equiv&1+\lambda^{d_2}.
\end{eqnarray}
The cascade form of the neutrino mass matrix requires the normal hierarchy of 
light neutrino mass spectrum, and mass eigenvalues are estimated as 
\begin{eqnarray}
  m_1&\simeq&\frac{\lambda^{2d}v_u^2}{6M}\equiv\bar{m}_1, \label{m1}\\
  m_2&\simeq&\left(3\lambda^{2d_1-x_1}+\frac{1}{3}\right)
  \frac{\lambda^{2d}v_u^2}{M}\equiv\bar{m}_2+2\bar{m_1}, \label{m2}\\
  m_3&\simeq&\left(2\lambda^{2d_2-x_2}+\frac{1}{2}\right)
  \frac{\lambda^{2d}v_u^2}{M}\equiv\bar{m}_3+3\bar{m_3}, 
  \label{m3}
\end{eqnarray}
including the leading order corrections of $\bar{m}_1$. 

Towards a profound understanding the hierarchical structure of the 
mass matrix and constraining on the cascade parameters, we rewrite the 
effective neutrino mass matrix~(\ref{neu-mass}) as 
\begin{eqnarray}
  M_\nu &\simeq& \frac{\lambda^{2d}v_u^2}{M}
  \left(
    \begin{array}{ccc}
      4  & -2 & -2 \\
      -2 & 1  & 1  \\
      -2 & 1  & 1
    \end{array}
  \right)+\frac{\lambda^{2(d_1+d)-x_1}v_u^2}{M}
  \left(
    \begin{array}{ccc}
      1 & 1 & 1 \\
      1 & 1 & 1 \\
      1 & 1 & 1
    \end{array}
  \right)\nonumber\\
  &&+\frac{\lambda^{2(d_2+d)-x_2}v_u^2}{M}
  \left(
    \begin{array}{ccc}
      0 & 0  & 0  \\
      0 & 1  & -1 \\
      0 & -1 & 1
    \end{array}
  \right)\nonumber\\
  &&+\frac{\lambda^{2d}v_u^2}{M}
  \left(
    \begin{array}{ccc}
      -4+\lambda^{2d_1}   & 2-\lambda^{d_1+d_2} & 2+\lambda^{d_1} \\
      2-\lambda^{d_1+d_2} & -1+\lambda^{2d_2}   & -1-\lambda^{d_2} \\
      2+\lambda^{d_1}     & -1-\lambda^{d_2}    & 0
    \end{array}
  \right)\nonumber\\
  &&+\frac{\lambda^{2(d_2+d)-x_2}v_u^2}{M}
  \left(
    \begin{array}{ccc}
      \lambda^{2(d_1-d_2)} & \lambda^{d_1-d_2} & -\lambda^{d_1-d_2} \\
      \lambda^{d_1-d_2}    & 0                 & 0                 \\
      \lambda^{d_1-d_2}    & 0                 & 0
    \end{array}
  \right).
  \label{neu-mass1}
\end{eqnarray}
In this mass matrix, if the terms in the first and second lines give 
dominant contribution, the tri-bimaximal mixing can be realized at the leading 
order. In order that the term in the third line of~(\ref{neu-mass1}) 
does not spoil the structures in the first line, $m_1\ll m_2,m_3$ is 
required. Therefore, the neutrino mass spectrum in the cascade model 
should be the normal mass hierarchy, as stated. Then it is well approximated 
that $m_2\simeq\sqrt{\Delta m_{21}^2}$ and $m_3\simeq\sqrt{|\Delta m_{31}^2|}$. 
Here one can obtain three constraints and/or relations 
for the cascade parameters. The first one comes from the hierarchy 
$m_1\ll m_2$. This means that $-2d_1+x_1\geq1$ for the parameters by 
using the~(\ref{m1}) and~(\ref{m2}). The second one is derived from 
the current experiments. Since the experimental data suggests that 
$r\equiv\sqrt{\Delta m_{21}^2}/\sqrt{|\Delta m_{31}^2|}\simeq0.18$, 
one obtains a relation among the cascade parameter as 
$2(d_1-d_2)-(x_1-x_2)=1$ or $2$ from~(\ref{m2}) and~(\ref{m3}), where 
the fact $\lambda\sim r$ is used. We also find that there exists a 
relation among the cascade parameters, light and heavy neutrino mass 
scales, that is, we can have a relation, 
$M\simeq\lambda^{2(d_2+d)-x_2}v_u^2/\sqrt{|\Delta m_{31}^2|}$, from 
$m_3\simeq\sqrt{|\Delta m_{31}^2|}$. Finally, we should consider the 
effects from the term in the last line of~(\ref{neu-mass1}). In order 
that the term does not spoil the democratic structure in the first 
line, the hierarchy $m_2\gg m_3\lambda^{d_1-d_2}$ is needed. This 
constraint can be expressed by the cascade parameters as 
$d_1-d_2-(x_1-x_2)\leq-1$. We conclude the above constraints and 
relations for the cascade parameters and physical quantities as 
\begin{eqnarray}
  \begin{array}{llll}
    (\mbox{i})   & m_1\ll m_2 & \Rightarrow & -2d_1+x_1\geq1, \\
    (\mbox{ii})  & m_2/m_3\simeq r\simeq 0.18 & \Rightarrow & 2(d_1-d_2)-(x_1-x_2)=1\mbox{ or }2, \\
    (\mbox{iii}) & m_3\simeq \sqrt{|\Delta m_{31}^2|} & \Rightarrow & M\simeq\lambda^{2(d_2+d)-x_2}v_u^2/\sqrt{|\Delta m_{31}^2|}, \\
    (\mbox{iv})  & m_2\gg m_3\lambda^{d_1-d_2} & \Rightarrow & d_1-d_2-(x_1-x_2)\leq-1.
  \end{array}
  \label{constraint1}
\end{eqnarray}
They restrict the neutrino Dirac mass matrix of the cascade form and the 
right-handed neutrino Majorana one of the diagonal form to textures shown in 
Tabs.~\ref{tab7} and~\ref{tab8}. 
\begin{table}
\begin{center}
\begin{tabular}{|c|c|c|c||c|c|}
\hline
$d_1$ & $d_2$ & $x_1$ & $x_2$ & $M_{\nu D}/(\lambda^dv_u)$ & $M_R/M$ \\
\hline
\hline
$3$ & $1$ & $7$ & $4$ & 
$
\left(
\begin{array}{ccc}
\lambda^3 & \lambda^3 & \lambda^3 \\
\lambda^3 & \lambda  & -\lambda \\
\lambda^3 & -\lambda & 1 
\end{array}
\right)
$ & 
$
\left(
\begin{array}{ccc}
\lambda^7 & 0 & 0 \\
0 & \lambda^4 & 0 \\
0 & 0 & 1 
\end{array}
\right)
$ \\
\hline
$3$ & $1$ & $8$ & $5$ & 
$
\left(
\begin{array}{ccc}
\lambda^3 & \lambda^3 & \lambda^3 \\
\lambda^3 & \lambda  & -\lambda \\
\lambda^3 & -\lambda & 1 
\end{array}
\right)
$ & 
$
\left(
\begin{array}{ccc}
\lambda^8 & 0 & 0 \\
0 & \lambda^5 & 0 \\
0 & 0 & 1 
\end{array}
\right)
$ \\
\hline
$3$ & $1$ & $\vdots$ & $\vdots$ & $\vdots$ & $\vdots$ \\
\hline
$4$ & $1$ & $9$ & $4$ & 
$
\left(
\begin{array}{ccc}
\lambda^4 & \lambda^4 & \lambda^4 \\
\lambda^4 & \lambda  & -\lambda \\
\lambda^4 & -\lambda & 1 
\end{array}
\right)
$ & 
$
\left(
\begin{array}{ccc}
\lambda^9 & 0 & 0 \\
0 & \lambda^4 & 0 \\
0 & 0 & 1 
\end{array}
\right)
$ \\
\hline
$4$ & $1$ & $\vdots$ & $\vdots$ & $\vdots$ & $\vdots$ \\
\hline
$4$ & $2$ & $9$ & $6$ & 
$
\left(
\begin{array}{ccc}
\lambda^4 & \lambda^4 & \lambda^4 \\
\lambda^4 & \lambda^2  & -\lambda^2 \\
\lambda^4 & -\lambda^2 & 1 
\end{array}
\right)
$ & 
$
\left(
\begin{array}{ccc}
\lambda^9 & 0 & 0 \\
0 & \lambda^6 & 0 \\
0 & 0 & 1 
\end{array}
\right)
$ \\
\hline
$\vdots$ & $\vdots$ & $\vdots$ & $\vdots$ & $\vdots$ & $\vdots$ \\
\hline
\end{tabular}
\end{center}
\caption{The textures of the neutrino Dirac mass matrix of the cascade form 
and the right-handed neutrino Majorana one of the diagonal form. 
The matrices are constrained 
by the experimental data of the neutrino masses with the 
condition $2(d_1-d_2)-(x_1-x_2)=1$.}
\label{tab7}
\end{table}
\begin{table}
\begin{center}
\begin{tabular}{|c|c|c|c||c|c|}
\hline
$d_1$ & $d_2$ & $x_1$ & $x_2$ & $M_{\nu D}/(\lambda^dv_u)$ & $M_R/M$ \\
\hline
\hline
$4$ & $1$ & $9$ & $5$ & 
$
\left(
\begin{array}{ccc}
\lambda^4 & \lambda^4 & \lambda^4 \\
\lambda^4 & \lambda  & -\lambda \\
\lambda^4 & -\lambda & 1 
\end{array}
\right)
$ & 
$
\left(
\begin{array}{ccc}
\lambda^9 & 0 & 0 \\
0 & \lambda^5 & 0 \\
0 & 0 & 1 
\end{array}
\right)
$ \\
\hline
$4$ & $1$ & $10$ & $6$ & 
$
\left(
\begin{array}{ccc}
\lambda^4 & \lambda^4 & \lambda^4 \\
\lambda^4 & \lambda  & -\lambda \\
\lambda^4 & -\lambda & 1 
\end{array}
\right)
$ & 
$
\left(
\begin{array}{ccc}
\lambda^{10} & 0 & 0 \\
0 & \lambda^6 & 0 \\
0 & 0 & 1 
\end{array}
\right)
$ \\
\hline
$4$ & $1$ & $\vdots$ & $\vdots$ & $\vdots$ & $\vdots$ \\
\hline
$5$ & $1$ & $11$ & $5$ & 
$
\left(
\begin{array}{ccc}
\lambda^5 & \lambda^5 & \lambda^5 \\
\lambda^5 & \lambda  & -\lambda \\
\lambda^5 & -\lambda & 1 
\end{array}
\right)
$ & 
$
\left(
\begin{array}{ccc}
\lambda^{11} & 0 & 0 \\
0 & \lambda^5 & 0 \\
0 & 0 & 1 
\end{array}
\right)
$ \\
\hline
$5$ & $1$ & $\vdots$ & $\vdots$ & $\vdots$ & $\vdots$ \\
\hline
$5$ & $2$ & $11$ & $7$ & 
$
\left(
\begin{array}{ccc}
\lambda^5 & \lambda^5 & \lambda^5 \\
\lambda^5 & \lambda^2  & -\lambda^2 \\
\lambda^5 & -\lambda^2 & 1 
\end{array}
\right)
$ & 
$
\left(
\begin{array}{ccc}
\lambda^{11} & 0 & 0 \\
0 & \lambda^7 & 0 \\
0 & 0 & 1 
\end{array}
\right)
$ \\
\hline
$\vdots$ & $\vdots$ & $\vdots$ & $\vdots$ & $\vdots$ & $\vdots$ \\
\hline
\end{tabular}
\end{center}
\caption{The textures of the neutrino Dirac mass matrix of the cascade form 
and the right-handed neutrino Majorana one of the diagonal form. 
The matrices are constrained 
by the experimental data of the neutrino masses with the 
condition $2(d_1-d_2)-(x_1-x_2)=2$.}
\label{tab8}
\end{table}
It is seen that the minimal model for the neutrino mass matrices is described 
by $(d,d_1,d_2,x_1,x_2)=(0,3,1,7,4)$ given in Tab.~\ref{tab7}. In this case, 
mass scale of the heaviest right-handed neutrino should be the order of 
the unification scale, namely $(M_1,M_2,M_3)\sim(10^{12},10^{14},10^{16})$ GeV. 

Let us comment on the mixing angles predicted by this cascade 
model. Even if the right-handed neutrino mass matrix is diagonal, the 
mixing angles deviate from the exact tri-bimaximal pattern. We discuss 
about these deviations predicted from the cascade model. The neutrino 
mass matrix after operating the tri-bimaximal mixing 
matrix~(\ref{neu-mass-diag}) can be rewritten as, 
\begin{eqnarray}
\mathcal{M}_0\simeq
    \begin{pmatrix}
      \bar{m}_1+\frac{\lambda^{2(d_1-d_2)}}{3}\bar{m}_3 & -\sqrt{2}\bar{m}_1+\frac{\sqrt{2}\lambda^{2(d_1-d_2)}}{6}\bar{m}_3 & -\sqrt{3}\bar{m}_1-\frac{\lambda^{d_1-d_2}}{\sqrt{3}}\bar{m}_3 \\
      -\sqrt{2}\bar{m}_1+\frac{\sqrt{2}\lambda^{2(d_1-d_2)}}{6}\bar{m}_3 & \bar{m}_2+2\bar{m}_1+\frac{\lambda^{2(d_1-d_2)}}{6}\bar{m}_3 & \sqrt{6}\bar{m}_1-\frac{\lambda^{d_1-d_2}}{\sqrt{6}}\bar{m}_3 \\
      -\sqrt{3}\bar{m}_1-\frac{\lambda^{d_1-d_2}}{\sqrt{3}}\bar{m}_3 & \sqrt{6}\bar{m}_1-\frac{\lambda^{d_1-d_2}}{\sqrt{6}}\bar{m}_3 & \bar{m}_3+3\bar{m}_1
    \end{pmatrix},
  \label{neu-mass-diag1}
\end{eqnarray}
up to leading order in each term of~(\ref{neu-mass-diag}). Note that 
if the cascade model realizes the exact tri-bimaximal mixing, this mass 
matrix should be diagonal. However, finite off-diagonal elements give 
deviations from the tri-bimaximal mixing. Let us estimate these 
deviations. The nearly diagonal neutrino mass 
matrix~(\ref{neu-mass-diag1}) can be diagonal by the following mixing 
matrix up to the next leading order, 
\begin{eqnarray}
  V^{(1)}\simeq\left(
    \begin{array}{ccc}
      1 & \theta_{12}^{(1)} & \theta_{13}^{(1)} \\
      -\theta_{12}^{(1)} & 1 & \theta_{23}^{(1)} \\
      -\theta_{13}^{(1)} & -\theta_{23}^{(1)} & 1 
    \end{array}
  \right),
\end{eqnarray} 
where 
\begin{eqnarray}
  \theta_{12}^{(1)}&\simeq&-\frac{\sqrt{2}\bar{m}_1}{\bar{m}_2}, \\
  \theta_{23}^{(1)}&\simeq&\frac{\sqrt{6}\bar{m}_1}{\bar{m}_3-\bar{m}_2}-\frac{\lambda^{d_1-d_2}\bar{m}_3}{\sqrt{6}(\bar{m}_3-\bar{m}_2)}, \\
  \theta_{13}^{(1)}&\simeq&-\frac{\sqrt{3}\bar{m}_1}{\bar{m}_3}-\frac{\lambda^{d_1-d_2}}{\sqrt{3}}. 
\end{eqnarray}
Therefore, the resultant PMNS matrix including these collection from the 
cascade structure (here we dropped correction from charged lepton sector), 
\begin{eqnarray}
  V_{\mbox{{\scriptsize PMNS}}}&\simeq&V_{\mbox{{\scriptsize TB}}}V^{(1)}P_M \\
  &=&\left(
    \begin{array}{ccc}
      2/\sqrt{6}  & 1/\sqrt{3} & 0           \\
      -1/\sqrt{6} & 1/\sqrt{3} & -1/\sqrt{2} \\
      -1/\sqrt{6} & 1/\sqrt{3} & 1/\sqrt{2}
    \end{array}
  \right)\left(
    \begin{array}{ccc}
      1 & \theta_{12}^{(1)} & \theta_{13}^{(1)} \\
      -\theta_{12}^{(1)} & 1 & \theta_{23}^{(1)} \\
      -\theta_{13}^{(1)} & -\theta_{23}^{(1)} & 1 
    \end{array}
  \right)P_M,
\end{eqnarray}
gives 
\begin{eqnarray}
  \sin^2\theta_{12} &\simeq& \left|\frac{1}{\sqrt{3}}
    +\frac{2}{\sqrt{6}}\theta_{12}^{(1)}\right|^2 
  \label{cor12} \\
  &\simeq& \left|\frac{1}{\sqrt{3}}
    -\frac{2}{\sqrt{3}}\frac{\bar{m}_1}{\bar{m}_2}
  \right|^2, \\
  \sin^2\theta_{23} &\simeq& \left|-\frac{1}{\sqrt{2}}
    -\frac{1}{\sqrt{6}}\theta_{13}^{(1)}
    +\frac{1}{\sqrt{3}}\theta_{23}^{(1)}\right|^2 \label{cor23} \\
  &\simeq& \left|-\frac{1}{\sqrt{2}}
    +\frac{1}{\sqrt{2}}
    \frac{\bar{m}_1(3\bar{m}_3-\bar{m}_2)}
    {\bar{m}_3(\bar{m}_3-\bar{m}_2)}
    -\frac{\lambda^{d_1-d_2}}{3\sqrt{2}}
    \frac{\bar{m}_2}{\bar{m}_3-\bar{m}_2}\right|^2, \\
  \sin^2\theta_{13} &\simeq& \left|\frac{2}{\sqrt{6}}\theta_{13}^{(1)}
    +\frac{1}{\sqrt{3}}\theta_{23}^{(1)}\right|^2 \label{cor13} \\
  &\simeq& \left|-\frac{\lambda^{d_1-d_2}}{\sqrt{2}}
    \frac{\bar{m}_3-\frac{2}{3}\bar{m}_2}
    {\bar{m}_3-\bar{m}_2}
    +\frac{\sqrt{2}\bar{m}_1\bar{m}_2}
    {\bar{m}_3(\bar{m}_3-\bar{m}_2)}\right|^2,
\end{eqnarray}
where $P_M$ is a diagonal phase matrix. It is seen that $\bar{m}_2$ and 
$\bar{m}_3$ are well approximated by $\sqrt{\Delta m_{21}^2}$ and 
$\sqrt{|\Delta m_{31}^2|}$, respectively, if $\bar{m}_1$ is sufficiently tiny. 

%%%%%%%%%%%%%%%%%%%%%%%%%%%%%%%%%%%%%%%
\subsubsection{Non-diagonal $M_R$ case}
%%%%%%%%%%%%%%%%%%%%%%%%%%%%%%%%%%%%%%%

Next, let us consider a case of non-diagonal $M_R$, which is 
generally allowed in the context of the cascade textures. Especially, 
corrections from a non-diagonal $M_R$ to the tri-bimaximal mixing angles are 
estimated. Some constraints for the corrections, equivalently the 
structure of $M_R$, are also presented. 

The neutrino Dirac mass matrix is taken as the cascade form given 
in~(\ref{Dirac}). We define the diagonalized mass matrix of the 
right-handed neutrino, $D_R$, as 
\begin{eqnarray}
  D_R\equiv U_{\nu R}^TM_RU_{\nu R}\equiv
  \left(
    \begin{array}{ccc}
      \lambda^{x_1} & 0             & 0 \\
      0             & \lambda^{x_2} & 0 \\
      0             & 0             & 1
    \end{array}
  \right)M \hspace{3mm}\mbox{ with }\hspace{3mm}x_1\geq x_2\geq0,\label{DR}
\end{eqnarray}
where $M_R$ is non-diagonal mass matrix for the right-handed neutrinos 
but mixing among each generation is assumed to be small in order to 
conserve the tri-bimaximal mixing at the leading order. If the mixing angles 
among each generation of the right-handed neutrino are enough small, the 
$U_{\nu  R}$ can be written as, 
\begin{eqnarray}
  U_{\nu R}\simeq 
  \left(
    \begin{array}{ccc}
      1              & \theta_{R,12}  & \theta_{R,13} \\
      -\theta_{R,12} & 1              & \theta_{R,23} \\
      -\theta_{R,13} & -\theta_{R,23} & 1
    \end{array}
  \right)
\equiv\left(
    \begin{array}{ccc}
      1                 & \lambda^{q_{12}}  & \lambda^{q_{13}} \\
      -\lambda^{q_{12}} & 1                 & \lambda^{q_{23}} \\
      -\lambda^{q_{13}} & -\lambda^{q_{23}} & 1
    \end{array}
%  \right) \mbox{ with }q_{12},q_{23},q_{13}\geq1,\label{UnuR}
  \right) \mbox{ with }q_{ij}\geq1,\label{UnuR}
\end{eqnarray}
up to the first order of $\theta_{R,ij}$ $(i,j=1\sim3)$. 

After the seesaw mechanism, one obtains the Majorana mass matrix of light 
neutrinos in low-energy theory as, 
\begin{eqnarray}
  M_\nu\simeq M_{\nu D}^TM_R^{-1}M_{\nu D}
  \simeq M_{\nu D}^TU_{\nu R}D_R^{-1}U_{\nu R}^TM_{\nu D}.
\end{eqnarray}
First, we write down $M_R^{-1}$ as 
\begin{eqnarray}
  M_R^{-1}&\simeq&U_{\nu R}D_R^{-1}U_{\nu R}^T \nonumber\\
  &=&\left(
    \begin{array}{ccc}
      h_1+\theta_{R,12}^2h_2+\theta_{R,13}^2h_3 & \theta_{R,12}h_{21}+\theta_{R,23}\theta_{R,13}h_3 & \theta_{R,13}h_{31}-\theta_{R,12}\theta_{R,23}h_2 \\
      \theta_{R,12}h_{21}+\theta_{R,23}\theta_{R,13}h_3 & h_2+\theta_{R,12}^2h_1+\theta_{R,23}^2h_3 & \theta_{R,23}h_{32}+\theta_{R,12}\theta_{R,13}h_1 \\
      \theta_{R,13}h_{31}-\theta_{R,12}\theta_{R,23}h_2 & \theta_{R,23}h_{32}+\theta_{R,12}\theta_{R,13}h_1 & h_3+\theta_{R,13}^2h_1+\theta_{R,23}^2h_2
    \end{array}
  \right),\nonumber\\
\end{eqnarray}
where $h_1\equiv(\lambda^{x_1}M)^{-1}$, $h_2\equiv(\lambda^{x_2}M)^{-1}$, 
$h_3\equiv M^{-1}$, and $h_{ij}\equiv h_i-h_j$. After the seesaw mechanism, the 
effective mass matrix of the light neutrinos is given by 
\begin{eqnarray}
  M_\nu&\simeq&M_{\nu D}^TM_R^{-1}M_{\nu D}\nonumber\\
  &\simeq&\left[\lambda^{2d_1}(M_R^{-1})_{11}\left(
      \begin{array}{ccc}
        1 & 1 & 1 \\
        1 & 1 & 1 \\
        1 & 1 & 1
      \end{array}\right)+\lambda^{2d_2}(M_R^{-1})_{22}\left(
      \begin{array}{ccc}
        0 & 0  & 0  \\
        0 & 1  & -1 \\
        0 & -1 & 1
      \end{array}\right)\right.\nonumber\\
  &&\phantom{\Bigg[}+\lambda^{2d_2}(M_R^{-1})_{22}\left(
    \begin{array}{ccc}
      \lambda^{2(d_1-d_2)} & \lambda^{d_1-d_2} & \lambda^{d_1-d_2} \\
      \lambda^{d_1-d_2} & 0 & 0 \\
      \lambda^{d_1-d_2} & 0 & 0
    \end{array}\right)\nonumber\\
  &&\phantom{\Bigg[}+(M_R^{-1})_{33}\left(
    \begin{array}{ccc}
      \lambda^{2d_1}     & -\lambda^{d_1+d_2} & \lambda^{d_1}  \\
      -\lambda^{d_1+d_2} & \lambda^{d_1+d_2}  & -\lambda^{d_2} \\
      \lambda^{d_1}      & -\lambda^{d_2}     & 1
    \end{array}\right)\nonumber\\
  &&\phantom{\Bigg[}+(M_R^{-1})_{23}\left(
    \begin{array}{ccc}
      2\lambda^{2d_1} & 0 & \lambda^{d_1}(1-\lambda^{d_2}) \\
      0 & -2\lambda^{2d_2} & \lambda^{d_2}(1+\lambda^{d_2}) \\
      \lambda^{d_1}(1-\lambda^{d_2}) & \lambda^{d_2}(1+\lambda^{d_2})     & -2\lambda^{d_2}
    \end{array}\right)\nonumber\\
  &&\phantom{\Bigg[}+\lambda^{d_1}(M_R^{-1})_{12}\left(
    \begin{array}{ccc}
      2 & \lambda^{d_1}+\lambda^{d_2} & \lambda^{d_1}-\lambda^{d_2} \\
      \lambda^{d_1}+\lambda^{d_2} & 2\lambda^{d_2} & 0 \\
      \lambda^{d_1}-\lambda^{d_2} & 0 & -2\lambda^{d_2}
    \end{array}\right)\nonumber\\
  &&\phantom{\Bigg[}\left.+\lambda^{d_1}(M_R^{-1})_{13}\left(
      \begin{array}{ccc}
        2\lambda^{d_1} & \lambda^{d_1}-\lambda^{d_2} & \lambda^{d_1}+1 \\
        \lambda^{d_1}-\lambda^{d_2} & -2\lambda^{d_2} & 1-\lambda^{d_2} \\
        \lambda^{d_1}+1 & 1-\lambda^{d_2} & 2
      \end{array}\right)\right]\lambda^{2d}v_u^2.\label{neu-mass2}
\end{eqnarray}
Note that if the $M_R$ is diagonal, which means $\theta_{R,ij}=0$, 
$(M_R^{-1})_{kl}$ $(k\neq l)$ are vanishing, and thus, the neutrino mass 
matrix~(\ref{neu-mass2}) results in~(\ref{neu-mass}). 

Operating the $V_{\mbox{{\scriptsize TB}}}$ to $M_\nu$ as 
$V_{\mbox{{\scriptsize TB}}}^TM_\nu V_{\mbox{{\scriptsize TB}}}$, the neutrino 
mass matrix becomes 
\begin{eqnarray}
  \mathcal{M}&\equiv&V_{\mbox{{\scriptsize TB}}}^TM_\nu 
  V_{\mbox{{\scriptsize TB}}}\nonumber\\
  &\simeq&\left[3\lambda^{2d_1}(M_R^{-1})_{11}\left(
      \begin{array}{ccc}
        0 & 0 & 0 \\
        0 & 1 & 0 \\
        0 & 0 & 0
      \end{array}\right)+2\lambda^{2d_2}(M_R^{-1})_{22}\left(
      \begin{array}{ccc}
        0 & 0 & 0 \\
        0 & 0 & 0 \\
        0 & 0 & 1
      \end{array}\right)\right.\nonumber\\
  &&\phantom{\Bigg[}+\frac{\lambda^{2d_2}(M_R^{-1})_{22}}{3}\left(
    \begin{array}{ccc}
      2\lambda^{2(d_1-d_2)} & \sqrt{2}\lambda^{2(d_1-d_2)} & 
      -2\sqrt{3}\lambda^{d_1-d_2} \\
      \sqrt{2}\lambda^{2(d_1-d_2)} & \lambda^{2(d_1-d_2)} & 
      -\sqrt{6}\lambda^{d_1-d_2} \\
      -2\sqrt{3}\lambda^{d_1-d_2} & -\sqrt{6}\lambda^{d_1-d_2} & 0
    \end{array}\right)\nonumber\\
  &&\phantom{\Bigg[}+\frac{(M_R^{-1})_{33}}{6}\left(
    \begin{array}{ccc}
      c_1^2           & -\sqrt{2}c_1c_2 & -\sqrt{3}c_1c_+ \\
      -\sqrt{2}c_1c_2 & 2c_2^2          & \sqrt{6}c_2c_+  \\
      -\sqrt{3}c_1c_+ & \sqrt{6}c_2c_+  & 3c_+^2
    \end{array}\right)\nonumber\\
  &&\phantom{\Bigg[}+\frac{(M_R^{-1})_{23}}{3\sqrt{2}}\left(
    \begin{array}{ccc}
      -2\sqrt{2}c_1\lambda^{d_1} & c_3\lambda^{d_1} & 
      \sqrt{6}c_-(\lambda^{d_2}+\lambda^{d_1}) \\
      c_3\lambda^{d_1} & 2\sqrt{2}c_2\lambda^{d_1} & 
      -\sqrt{3}c_-(2\lambda^{d_2}-\lambda^{d_1}) \\
      \sqrt{6}c_-(\lambda^{d_2}+\lambda^{d_1}) & -\sqrt{3}c_-(2\lambda^{d_2}-\lambda^{d_1}) & -6\sqrt{2}c_+\lambda^{d_2}
    \end{array}\right)\nonumber\\
  &&\phantom{\Bigg[}+\lambda^{d_1}(M_R^{-1})_{12}\left(
    \begin{array}{ccc}
      0 & \sqrt{2}\lambda^{d_1} & 0 \\
      \sqrt{2}\lambda^{d_1} & 2\lambda^{d_1} & -\sqrt{6}\lambda^{d_2} \\
      0 & -\sqrt{6}\lambda^{d_2} & 0
    \end{array}\right)\nonumber\\
  &&\phantom{\Bigg[}\left.
    +\frac{\lambda^{d_1}(M_R^{-1})_{13}}{\sqrt{2}}\left(
      \begin{array}{ccc}
        0 & -c_1 & 0 \\
        -c_1 & 2\sqrt{2}c_2 & \sqrt{3}c_+ \\
        0 & \sqrt{3}c_+ & 0
      \end{array}\right)\right]\lambda^{2d}v_u^2,
  \label{off-neu-mm}
\end{eqnarray}
where 
\begin{eqnarray}
  c_3&\equiv&1-\lambda^{d_2}+4\lambda^{d_1}, \\
  c_-&\equiv&1-\lambda^{d_2}.
\end{eqnarray}
This mass matrix can be written as 
\begin{eqnarray}
  \mathcal{M}=\mathcal{M}_0+\mathcal{M}_{\mbox{{\scriptsize off}}} 
  \equiv\mathcal{M}_0+\left(
    \begin{array}{ccc}
      m_1^R & m_{12}^R & m_{13}^R \\
      m_{12}^R & m_2^R & m_{23}^R \\
      m_{13}^R & m_{23}^R & m_3^R
    \end{array}
  \right),
\end{eqnarray}
where $\mathcal{M}_0$ comes from the diagonal elements of $M_R$, which 
is given by \eqref{neu-mass-diag1}. The matrix 
$\mathcal{M}_{\mbox{{\scriptsize off}}}$ includes effects from 
off-diagonal elements of $M_R$, whose elements are described by 
$\theta_{R,ij}$. If the mixing angles among the right-handed neutrinos are 
small, the neutrino mass matrix $\mathcal{M}$ should be almost 
diagonal. This means that the off-diagonal elements of $\mathcal{M}$ give 
collections to the tri-bimaximal generation mixing. These collections 
should be enough small to explain the MNS matrix, because the nearly 
tri-bimaximal structure can be realized by the neutrino Dirac mass matrix 
of the cascade form with the seesaw mechanism in the case of diagonal $M_R$. 
If a structure of $M_R$ leads to relatively large collections, then 
unnatural cancellations are needed to predict experimentally favored 
generation mixing angles of the lepton sector in the context of cascade 
neutrino Dirac mass matrix. Therefore, we focus on 
a case that the collections from the off-diagonal elements of $M_R$ 
are enough small not to spoil the nearly tri-bimaximal mixing induced 
from the cascade neutrino Dirac mass matrix and discuss about the 
magnitude of collections in the case. This means that the structure of 
resultant neutrino mass matrix given in \eqref{off-neu-mm} should not 
be drastically different from the \eqref{neu-mass-diag1} in the 
diagonal $M_R$ case. Thus the magnitude of neutrino mass 
eigenvalues given in \eqref{m1}$\sim$\eqref{m3} and the constraints 
\eqref{constraint1} must be held at the leading order even in the case 
of non-diagonal $M_R$. These discussions gives the following 
neutrino mass eigenvalues up to the next leading 
order,\footnote{Detailed discussions is given in the Appendix.} 
\begin{eqnarray}
  m_1&\simeq&\frac{\lambda^{2d}v_u^2}{6M}+m_1^R=\bar{m_1}+m_1^R, \\
  m_2&\simeq&\left(3\lambda^{2d_1-x_1}+\frac{1}{3}\right)
  \frac{\lambda^{2d}v_u^2}{M}+m_2^R=\bar{m}_2+m_2^R+2\bar{m_1}, \\
  m_3&\simeq&\left(2\lambda^{2d_2-x_2}+\frac{1}{2}\right)
  \frac{\lambda^{2d}v_u^2}{M}+m_3^R=\bar{m_3}+m_3^R+3\bar{m_3},   
\end{eqnarray}
where $m_i^R$ include effects from the off-diagonal element of $M_R$ 
described as 
\begin{eqnarray}
  m_1^R &\equiv& \frac{\lambda^{2d}v_u^2}{6M}\lambda^{-x_1}\theta_{R,23}^2, \\
  m_2^R &\equiv& -\frac{2\lambda^{2d}v_u^2}{3M}\lambda^{d_1}
  (3\lambda^{-x_1}\theta_{R,13}+\lambda^{-x_2}\theta_{R,23}^2), \\ 
  m_3^R &\equiv& \frac{\lambda^{2d}v_u^2}{2M}\lambda^{-x_1}
  (2\lambda^{d_2}\theta_{R,12}-\theta_{R,13})^2.
\end{eqnarray}
Typical textures of non-diagonal $M_R$ are given in Tabs.~\ref{tab9} 
and~\ref{tab10}.
\begin{table}
\begin{center}
\begin{tabular}{|c|c|c|c||c|c|}
\hline
$d_1$ & $d_2$ & $x_1$ & $x_2$ & $M_{\nu D}/(\lambda^dv_u)$ & $M_R/M$ \\
\hline
\hline
$3$ & $1$ & $7$ & $4$ & 
$
\left(
\begin{array}{ccc}
\lambda^3 & \lambda^3 & \lambda^3 \\
\lambda^3 & \lambda  & -\lambda \\
\lambda^3 & -\lambda & 1 
\end{array}
\right)
$ & 
$
\left(
\begin{array}{ccc}
\lambda^7 & \lambda^9 & \lambda^5 \\
\lambda^9 & \lambda^4 & \lambda^3 \\
\lambda^5 & \lambda^3 & 1 
\end{array}
\right)
$ \\
\hline
$3$ & $1$ & $8$ & $5$ & 
$
\left(
\begin{array}{ccc}
\lambda^3 & \lambda^3 & \lambda^3 \\
\lambda^3 & \lambda  & -\lambda \\
\lambda^3 & -\lambda & 1 
\end{array}
\right)
$ & 
$
\left(
\begin{array}{ccc}
\lambda^8 & \lambda^{11} & \lambda^6 \\
\lambda^{11} & \lambda^5 & \lambda^5 \\
\lambda^6 & \lambda^5 & 1 
\end{array}
\right)
$ \\
\hline
$3$ & $1$ & $\vdots$ & $\vdots$ & $\vdots$ & $\vdots$ \\
\hline
$4$ & $1$ & $9$ & $4$ & 
$
\left(
\begin{array}{ccc}
\lambda^4 & \lambda^4 & \lambda^4 \\
\lambda^4 & \lambda  & -\lambda \\
\lambda^4 & -\lambda & 1 
\end{array}
\right)
$ & 
$
\left(
\begin{array}{ccc}
\lambda^9 & \lambda^{10} & \lambda^6 \\
\lambda^{10} & \lambda^4 & \lambda^4 \\
\lambda^6 & \lambda^4 & 1 
\end{array}
\right)
$ \\
\hline
$4$ & $1$ & $\vdots$ & $\vdots$ & $\vdots$ & $\vdots$ \\
\hline
$4$ & $2$ & $9$ & $6$ & 
$
\left(
\begin{array}{ccc}
\lambda^4 & \lambda^4 & \lambda^4 \\
\lambda^4 & \lambda^2  & -\lambda^2 \\
\lambda^4 & -\lambda^2 & 1 
\end{array}
\right)
$ & 
$
\left(
\begin{array}{ccc}
\lambda^9 & \lambda^{11} & \lambda^5 \\
\lambda^{11} & \lambda^6 & \lambda^5 \\
\lambda^5 & \lambda^5 & 1 
\end{array}
\right)
$ \\ 
\hline
$\vdots$ & $\vdots$ & $\vdots$ & $\vdots$ & $\vdots$ & $\vdots$ \\
\hline
\end{tabular}
\end{center}
\caption{The textures of the neutrino Dirac mass matrix of the cascade form and 
the right-handed neutrino Majorana one constrained by the experimentally 
observed values of the neutrino masses with the condition 
$2(d_1-d_2)-(x_1-x_2)=1$.}
\label{tab9}
\end{table}
\begin{table}
\begin{center}
\begin{tabular}{|c|c|c|c||c|c|}
\hline
$d_1$ & $d_2$ & $x_1$ & $x_2$ & $M_{\nu D}/(\lambda^dv_u)$ & $M_R/M$ \\
\hline
\hline
$4$ & $1$ & $9$ & $5$ & 
$
\left(
\begin{array}{ccc}
\lambda^4 & \lambda^4 & \lambda^4 \\
\lambda^4 & \lambda  & -\lambda \\
\lambda^4 & -\lambda & 1 
\end{array}
\right)
$ & 
$
\left(
\begin{array}{ccc}
\lambda^9 & \lambda^{11} & \lambda^6 \\
\lambda^{11} & \lambda^5 & \lambda^5 \\
\lambda^6 & \lambda^5 & 1 
\end{array}
\right)
$\\
\hline
$4$ & $1$ & $10$ & $6$ & 
$
\left(
\begin{array}{ccc}
\lambda^4 & \lambda^4 & \lambda^4 \\
\lambda^4 & \lambda  & -\lambda \\
\lambda^4 & -\lambda & 1 
\end{array}
\right)
$ & 
$
\left(
\begin{array}{ccc}
\lambda^{10} & \lambda^{13} & \lambda^7 \\
\lambda^{13} & \lambda^6 & \lambda^6 \\
\lambda^7 & \lambda^6 & 1 
\end{array}
\right)
$ \\
\hline
$4$ & $1$ & $\vdots$ & $\vdots$ & $\vdots$ & $\vdots$ \\
\hline
$5$ & $1$ & $11$ & $5$ & 
$
\left(
\begin{array}{ccc}
\lambda^5 & \lambda^5 & \lambda^5 \\
\lambda^5 & \lambda  & -\lambda \\
\lambda^5 & -\lambda & 1 
\end{array}
\right)
$ & 
$
\left(
\begin{array}{ccc}
\lambda^{11} & \lambda^{12} & \lambda^7 \\
\lambda^{12} & \lambda^5 & \lambda^5 \\
\lambda^7 & \lambda^5 & 1 
\end{array}
\right)
$\\
\hline
$5$ & $1$ & $\vdots$ & $\vdots$ & $\vdots$ & $\vdots$ \\
\hline
$5$ & $2$ & $11$ & $7$ & 
$
\left(
\begin{array}{ccc}
\lambda^5 & \lambda^5 & \lambda^5 \\
\lambda^5 & \lambda^2  & -\lambda^2 \\
\lambda^5 & -\lambda^2 & 1 
\end{array}
\right)
$ & 
$
\left(
\begin{array}{ccc}
\lambda^{11} & \lambda^{13} & \lambda^7 \\
\lambda^{13} & \lambda^7 & \lambda^6 \\
\lambda^7 & \lambda^6 & 1 
\end{array}
\right)
$ \\
\hline
$\vdots$ & $\vdots$ & $\vdots$ & $\vdots$ & $\vdots$ & $\vdots$ \\
\hline
\end{tabular}
\end{center}
\caption{The textures of the neutrino Dirac mass matrix of the cascade form and 
the right-handed neutrino Majorana one constrained by the experimentally 
observed values of the neutrino masses with the condition 
$2(d_1-d_2)-(x_1-x_2)=2$.}
\label{tab10}
\end{table}
All the presented textures of $M_R$ preserve tri-bimaximal neutrino mixing 
at the leading order with relatively small numbers of $(d_1,d_2,x_1,x_2)$. 

The collections to the tri-bimaximal mixing are estimated in the 
perturbative method as in the diagonal $M_R$ case, 
\begin{eqnarray}
  \theta_{12}^{(1)}&\simeq&-\frac{\sqrt{2}\bar{m}_1+m_{12}^R}{\bar{m}_2+m_2^R}, \\
  \theta_{23}^{(1)}&\simeq&\frac{\sqrt{6}\bar{m}_1-\frac{\lambda^{d_1-d_2}}{\sqrt{6}}\bar{m}_3+m_{23}^R}{(\bar{m}_3+m_3^R)-(\bar{m}_2+m_2^R)}, \\
  \theta_{13}^{(1)}&\simeq&\frac{-\sqrt{3}\bar{m}_1-\frac{\lambda^{d_1-d_2}}{\sqrt{3}}\bar{m}_3+m_{13}^R}{\bar{m}_3+m_3^R}, 
\end{eqnarray}
where 
\begin{eqnarray}
  m_{12}^R &\simeq& -\frac{1}{6\sqrt{2}}
  [\lambda^{-2d_2}(\lambda^{d_1}
  +\theta_{R,23})\theta_{R,23}\bar{m}_3
  -2(2\theta_{R,12}-\lambda^{-d_1}\theta_{R,13})\bar{m}_2
  ], \\
  m_{23}^R &\simeq& \frac{1}{\sqrt{6}}[\theta_{R,23}\bar{m}_3+\lambda^{-d_1}(2\lambda^{d_2}\theta_{R,12}-\theta_{R,13})(1-\lambda^{-d_1}\theta_{R,13})\bar{m}_2], \\
  m_{13}^R &\simeq& -\frac{1}{4\sqrt{3}}\lambda^{-d_2}(2+\lambda^{-d_2}\theta_{R,23})
  \theta_{R,23}\bar{m}_3.
\end{eqnarray}
Finally, the collections to the PMNS mixing angles are 
\begin{eqnarray}
  \sin\theta_{12}
  &\simeq& \frac{1}{\sqrt{3}}
  +\frac{2}{\sqrt{6}}\frac{-\bar{m}_1+m_{12}^R}{\bar{m}_2+m_2^R}, \\
  \sin\theta_{23}
  &\simeq&-\frac{1}{\sqrt{2}}
  +\frac{1}{\sqrt{2}}
  \frac{\bar{m}_1[3(\bar{m}_3+m_3^R)-(\bar{m}_2+m_2^R)]}
  {(\bar{m}_3+m_3^R)[(\bar{m}_3+m_3^R)-(\bar{m}_2+m_2^R)]} 
  \nonumber \\
  &      & -\frac{\lambda^{d_1-d_2}}{3\sqrt{2}}
  \frac{\bar{m}_3(\bar{m}_2+m_2^R)}
  {(\bar{m}_3+m_3^R)[(\bar{m}_3+m_3^R)-(\bar{m}_2+m_2^R)}
  -\frac{1}{\sqrt{6}}\frac{m_{13}^R}{\bar{m}_3+m_3^R} \nonumber \\
  &      & +\frac{1}{\sqrt{3}}
  \frac{m_{23}^R}{(\bar{m}_3+m_3^R)-(\bar{m}_2+m_2^R)},\\
  \sin\theta_{13}
  &\simeq& -\frac{\lambda^{d_1-d_2}}{\sqrt{2}}
  \frac{\bar{m}_3\left[(\bar{m}_3+m_3^R)
      -\frac{2}{3}(\bar{m}_2+m_2^R)\right]}
  {(\bar{m}_3+m_3^R)[(\bar{m}_3+m_3^R)-(\bar{m}_2+m_2^R)]} 
  \nonumber \\
  &     & +\frac{\sqrt{2}\bar{m}_1(\bar{m}_2+m_2^R)}
  {(\bar{m}_3+m_3^R)[(\bar{m}_3+m_3^R)-(\bar{m}_2+m_2^R)]}
  +\frac{2}{\sqrt{6}}\frac{m_{13}^R}{\bar{m}_3+m_3^R} \nonumber \\
  &     & +\frac{1}{\sqrt{3}}
  \frac{m_{23}^R}{(\bar{m}_3+m_3^R)-(\bar{m}_2+m_2^R)}. 
\end{eqnarray}

%%%%%%%%%%%%%%%%%%%%%%%%%%%%%%%%%%%%%%%
\subsection{Charged lepton sector}\label{sec-clepton}
%%%%%%%%%%%%%%%%%%%%%%%%%%%%%%%%%%%%%%%

As mentioned above, we explore the possibility that the mass matrix of 
charged leptons has the H.C. form which is restricted by the GUT 
relation of $SU(5)$, $M_e\simeq M_d^T$. In this subsection, we study 
the corrections from the charged lepton sector to the lepton 
generation mixing angles. 

We take the charged lepton mass matrix as the following form, 
\begin{eqnarray}
  M_e\simeq 
  \left(
    \begin{array}{ccc}
      \epsilon_d & \delta_d    & \delta_d \\
      \delta_d   & -3\lambda_d & \lambda_d \\
      \delta_d   & \lambda_d   & 1
    \end{array}
  \right)\xi_dv_d,
\end{eqnarray}
Unlike the neutrino sector, the magnitudes of H.C. can be partially 
evaluated from the experimentally observed values of charged lepton 
masses and given by 
\begin{eqnarray}
  |\lambda_d| \simeq \frac{m_\mu}{3m_\tau},~~~%\simeq 1.99\times10^{-2}
  % \simeq  1.70\lambda^3, 
  |\delta_d|  \simeq \frac{3\sqrt{m_em_\mu}}{m_\tau}.%\simeq4.14\times10^{-3}
  % \simeq  1.56\lambda^4.
\end{eqnarray}
The generation mixing is expressed in terms of the cascade hierarchy 
parameter, $\lambda_d$ and $\delta_d$, as shown in 
Tab.~\ref{tab2}. Therefore, the corrections from the charged lepton 
sector are found to be generically small and the total lepton mixing 
angles are given at the first order of perturbation as 
\begin{eqnarray}
  \sin\theta_{12}
  &\simeq& \frac{1}{\sqrt{3}}
  +\frac{2}{\sqrt{6}}\frac{-\bar{m}_1+m_{12}^R}{\bar{m}_2+m_2^R}
  +\sqrt{\frac{3m_e}{m_\mu}}, \\
  \sin\theta_{23}
  &\simeq&-\frac{1}{\sqrt{2}}
  +\frac{1}{\sqrt{2}}
  \frac{\bar{m}_1[3(\bar{m}_3+m_3^R)-(\bar{m}_2+m_2^R)]}
  {(\bar{m}_3+m_3^R)[(\bar{m}_3+m_3^R)-(\bar{m}_2+m_2^R)]} 
  \nonumber \\
  &      & -\frac{\lambda^{d_1-d_2}}{3\sqrt{2}}
  \frac{\bar{m}_3(\bar{m}_2+m_2^R)}
  {(\bar{m}_3+m_3^R)[(\bar{m}_3+m_3^R)-(\bar{m}_2+m_2^R)}
  -\frac{1}{\sqrt{6}}\frac{m_{13}^R}{\bar{m}_3+m_3^R} \nonumber \\
  &      & +\frac{1}{\sqrt{3}}
  \frac{m_{23}^R}{(\bar{m}_3+m_3^R)-(\bar{m}_2+m_2^R)}
  +\frac{m_\mu}{3m_\tau},\\
  \sin\theta_{13}
  &\simeq& -\frac{\lambda^{d_1-d_2}}{\sqrt{2}}
  \frac{\bar{m}_3\left[(\bar{m}_3+m_3^R)
      -\frac{2}{3}(\bar{m}_2+m_2^R)\right]}
  {(\bar{m}_3+m_3^R)[(\bar{m}_3+m_3^R)-(\bar{m}_2+m_2^R)]} 
  \nonumber \\
  &     & +\frac{\sqrt{2}\bar{m}_1(\bar{m}_2+m_2^R)}
  {(\bar{m}_3+m_3^R)[(\bar{m}_3+m_3^R)-(\bar{m}_2+m_2^R)]}
  +\frac{2}{\sqrt{6}}\frac{m_{13}^R}{\bar{m}_3+m_3^R} \nonumber \\
  &     & +\frac{1}{\sqrt{3}}
  \frac{m_{23}^R}{(\bar{m}_3+m_3^R)-(\bar{m}_2+m_2^R)}
  +\frac{3}{\sqrt{2}}\sqrt{\frac{m_e}{m_\mu}}.
\end{eqnarray}
One can see the effects from the charged lepton sector of the 
H.C. form from these expressions. The tri-bimaximal solar neutrino 
mixing is little (about $4\%$ of $\sin^2\theta_{12}$) affected. For 
the atmospheric neutrino mixing, the charged lepton effect becomes 
$4\%$ of the tri-bimaximal atmospheric angle, 
$\sin^2\theta_{23}=1/2$. Finally, magnitude of effect is estimated 
as $0.02$ for the reactor neutrino angle, $\sin^2\theta_{13}$. 

%%%%%%%%%%%%%%%%%%%%%%%%%%%%%%
\subsection{Quark sector}
%%%%%%%%%%%%%%%%%%%%%%%%%%%%%%
We investigate the quark mass matrices in this subsection. It must be 
remembered that the mass matrix of the H.C. form is motivated for the 
mass spectra and mixing angles of quark sector. The cascading mass matrices 
of down- and up-type quarks are given in~\eqref{cas-mu} and~\eqref{cas-md}. 
The mixing matrices for the down and up sector are roughly estimated as 
\begin{eqnarray}
  V_d =
  \left(
    \begin{array}{lll}
      \mathcal{O}(1) & \mathcal{O}(\lambda) & \mathcal{O}(\lambda^3) \\
      \mathcal{O}(\lambda) & \mathcal{O}(1) & \mathcal{O}(\lambda^2) \\
      \mathcal{O}(\lambda^3) & \mathcal{O}(\lambda^2) & \mathcal{O}(1)
    \end{array}
  \right),~~~ 
  V_u =
  \left(
    \begin{array}{lll}
      \mathcal{O}(1) & \mathcal{O}(\lambda^4) & \mathcal{O}(\lambda^8) \\
      \mathcal{O}(\lambda^4) & \mathcal{O}(1) & \mathcal{O}(\lambda^4) \\
      \mathcal{O}(\lambda^8) & \mathcal{O}(\lambda^4) & \mathcal{O}(1)
    \end{array}
  \right),
\end{eqnarray}
where the mass matrix for the up quarks is assumed to be a symmetric 
matrix. It is easily seen from the structure of $V_d$ that the 
experimentally observed values of CKM matrix can be realized at the 
leading order. The collections from the $V_u$ are generally small, 
which are estimated as 
\begin{eqnarray}
  |V_{td}| &\simeq& |(V_d)_{31}+(V_d)_{21}(V_u^\dagger)_{23}|
  \simeq  |\lambda^3(1+\lambda^2)|, \\
  |V_{cb}| &\simeq& |(V_d)_{23}+(V_u^\dagger)_{32}|\simeq|\lambda(1+\lambda^2)|, \\ 
  |V_{ts}| &\simeq& |(V_d)_{32}+(V_u)_{23}^\dagger|
  \simeq  |\lambda^2(1+\lambda^2)|,
\end{eqnarray}
up to order $\mathcal{O}(\lambda^2)$ of the dominant term, and 
\begin{eqnarray}
  |V_{us}| &\simeq& |(V_d)_{12}+(V_u^\dagger)_{21}|\simeq|\lambda(1+\lambda^3)|, \\
  |V_{ub}| &\simeq& |(V_d)_{13}+(V_u^\dagger)_{21}(V_d^\dagger)_{23}|
  \simeq  |\lambda^3(1+\lambda^3)|, \\
  |V_{cd}| &\simeq& |(V_d)_{21}+(V_u^\dagger)_{12}|\simeq|\lambda(1+\lambda^3)|, 
\end{eqnarray}
up to order $\mathcal{O}(\lambda^3)$ of the leading term. Collections to other 
elements are negligibly small. Detailed 
numerical calculations are given in the next section. 

%%%%%%%%%%%%%%%%%%%%%%%%%%%%%%%
\section{Related Phenomenology}
%%%%%%%%%%%%%%%%%%%%%%%%%%%%%%%

In this section, we numerically investigate related phenomenologies 
based on the above analyses of cascade textures for the quark and 
lepton sectors: the generation mixing angles, the lepton flavor violation, 
and the baryon asymmetry of the Universe via thermal leptogenesis. 

%%%%%%%%%%%%%%%%%%%%%%%%%%%%%%%
\subsection{Generation mixing angles}
\label{sec-genmix}
%%%%%%%%%%%%%%%%%%%%%%%%%%%%%%%

Let us start to examine numerical analyses of the generation mixing 
of the quark and lepton sectors predicted from the cascade model. In 
these analyses, we focus on two typical types of minimal texture for 
the neutrino Dirac and right-handed Majorana neutrino mass matrices 
given in Tabs.~\ref{tab9} and~\ref{tab10}, that is, 
\begin{eqnarray}\label{m1n}
  \mbox{Model I : }~~~
  M_{\nu D}\simeq 
  \left(
    \begin{array}{ccc}
      \lambda^3 & \lambda^3 & \lambda^3 \\
      \lambda^3 & \lambda  & -\lambda \\
      \lambda^3 & -\lambda & 1 
    \end{array}
  \right)\lambda^dv_u,~~~
  M_R\simeq 
  \left(
    \begin{array}{ccc}
      \lambda^7 & \lambda^9 & \lambda^5 \\
      \lambda^9 & \lambda^4 & \lambda^3 \\
      \lambda^5 & \lambda^3 & 1 
    \end{array}
  \right)M,
\end{eqnarray}
for the case of the condition $2(d_1-d_2)-(x_1-x_2)=1$, and 
\begin{eqnarray}\label{m2n}
  \mbox{Model II : }~~~
  M_{\nu D}\simeq 
  \left(
    \begin{array}{ccc}
      \lambda^4 & \lambda^4 & \lambda^4 \\
      \lambda^4 & \lambda  & -\lambda \\
      \lambda^4 & -\lambda & 1 
    \end{array}
  \right)\lambda^dv_u,~~~
  M_R\simeq 
  \left(
    \begin{array}{ccc}
      \lambda^9 & \lambda^{11} & \lambda^6 \\
      \lambda^{11} & \lambda^5 & \lambda^5 \\
      \lambda^6 & \lambda^5 & 1 
    \end{array}
  \right)M,
\end{eqnarray}
for the case of the condition $2(d_1-d_2)-(x_1-x_2)=2$. In both models, the 
following charged lepton, up and down quark mass matrices are adopted: 
\begin{eqnarray}
  M_e\simeq 
  \left(
    \begin{array}{ccc}
      \lambda^{k_d+3} & \lambda^3    & \lambda^3 \\
      \lambda^3   & -3\lambda^2 & \lambda^2 \\
      \lambda^3   & \lambda^2   & 1
    \end{array}
  \right)\xi_dv_d,
\end{eqnarray}
and 
\begin{eqnarray}\label{mqneum}
  M_u\simeq 
  \left(
    \begin{array}{ccc}
      \lambda^{k_u+6} & \lambda^6 & \lambda^6 \\
      \lambda^6       & \lambda^4 & \lambda^4 \\
      \lambda^6       & \lambda^4 & 1
    \end{array}
  \right)v_u,~~~
  M_d\simeq 
  \left(
    \begin{array}{ccc}
      \lambda^{k_d+3} & \lambda^3 & \lambda^3 \\
      \lambda^3       & \lambda^2 & \lambda^2 \\
      \lambda^3       & \lambda^2 & 1
    \end{array}
  \right)\xi_dv_d.
\end{eqnarray}

Using the above mass hierarchies, we can numerically examine 
predictions of PMNS angles, and can compare the predictions of model I 
and model II. The analysis is carried out with the following 
procedure. At first, we restrict structure in $M_u$ and $M_d$ with the 
experimental constraints on quark masses and the CKM matrix. Then the 
GUT relation between charged lepton and down-type quark provides a 
constrained structure in $M_e$. For the neutrino sector, different 
contributions to PMNS angles are obtained for each model. Combining 
contributions from $M_e$ and neutrino sector, predictions of PMNS 
angles are obtained. 

Note that \eqref{m1n} and~\eqref{mqneum} include 
ambiguity from ${\cal O}(1)$ coefficients of each element of  
matrices. In the numerical estimation, the definite form of the 
mass matrices are used. In the neutrino sector, the mass matrices are 
defined as 
\begin{eqnarray}\label{m1n_wc}
  % \mbox{Model I : }~~~
  M_{\nu D}=
  \left(
    \begin{array}{ccc}
      c_\nu\lambda^3 & c_\nu\lambda^3 & c_\nu\lambda^3 \\
      c_\nu\lambda^3 & b_\nu\lambda  & -b_\nu\lambda \\
      c_\nu\lambda^3 & -b_\nu\lambda & a_\nu
    \end{array}
  \right)\lambda^dv_u,~~~
  M_R=
  \left(
    \begin{array}{ccc}
      f_R\lambda^7 & e_R\lambda^9 & d_R\lambda^5 \\
      e_R\lambda^9 & c_R\lambda^4 & b_R\lambda^3 \\
      d_R\lambda^5 & b_R\lambda^3 & a_R
    \end{array}
  \right)M,
\end{eqnarray}
for model I, and 
\begin{eqnarray}\label{m2n_wc}
  % \mbox{Model II : }~~~
  M_{\nu D}=
  \left(
    \begin{array}{ccc}
      c_\nu\lambda^4 & c_\nu\lambda^4 & c_\nu\lambda^4 \\
      c_\nu\lambda^4 &  b_\nu\lambda  & -b_\nu\lambda \\
      c_\nu\lambda^4 & -b_\nu\lambda & a_\nu
    \end{array}
  \right)\lambda^dv_u,~~~
  M_R=
  \left(
    \begin{array}{ccc}
      f_R\lambda^9 & e_R\lambda^{11} & d_R\lambda^6 \\
      e_R\lambda^{11} & c_R\lambda^5 & b_R\lambda^5 \\
      d_R\lambda^6 & b_R\lambda^5 & a_R
    \end{array}
  \right)M,
\end{eqnarray}
for model II. For the quark and charged lepton mass matrices, we take 
\begin{eqnarray}\label{mqneum_wc}
  M_u=
  \left(
    \begin{array}{ccc}
      0 & e_u\lambda^6 & d_u\lambda^6 \\
      e_u\lambda^6       & c_u\lambda^4 & b_u\lambda^4 \\
      d_u\lambda^6       & b_u\lambda^4 & a_u
    \end{array}
  \right)v_u,~~~
  M_d=
  \left(
    \begin{array}{ccc}
      0 & e_d\lambda^3 & d_d\lambda^3 \\
      e_d\lambda^3       & c_d\lambda^2 & b_d\lambda^2 \\
      d_d\lambda^3       & b_d\lambda^2 & a_d
    \end{array}
  \right)\xi_dv_d.
\end{eqnarray}
and 
\begin{eqnarray}
  M_e=
  \left(
    \begin{array}{ccc}
      0 & e_d\lambda^3    & d_d\lambda^3 \\
      e_d\lambda^3   & -3c_d\lambda^2 & b_d\lambda^2 \\
      d_d\lambda^3   & b_d\lambda^2   & a_d
    \end{array}
  \right)\xi_dv_d. 
\end{eqnarray}
In the matrices, the coefficients represented by $a,\cdots,f$ with subscripts 
are taken as complex numbers whose absolute values are constrained as 
0.4$\sim$1.4. The $(1,1)$ elements of $M_{u,d,e}$ are taken as 
zero. The limitation on the mass matrices is not essential, that is, the 
following results are little affected by the $(1,1)$ elements that 
satisfy the condition shown in section~\ref{sec-op}. 

%%%%%%%%%% FIGURE.1 
\begin{figure}[t]
\begin{center}
  \includegraphics[width=7cm,clip]{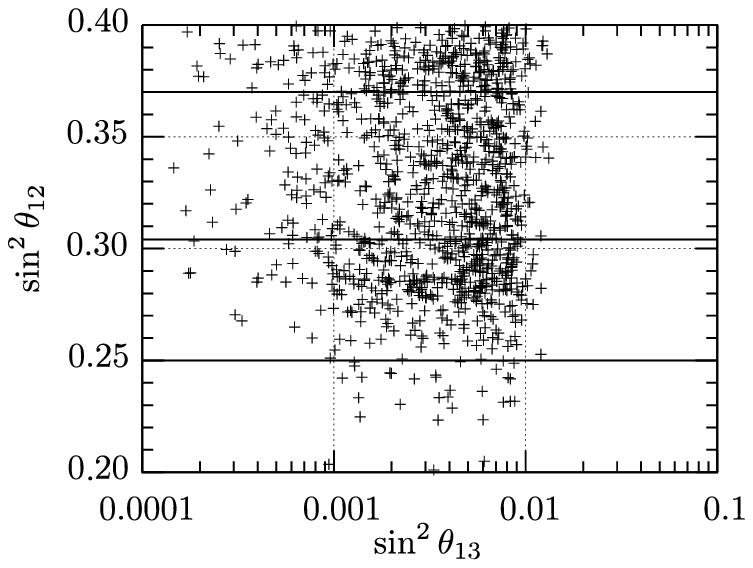}\hspace{1cm}
  \includegraphics[width=7cm,clip]{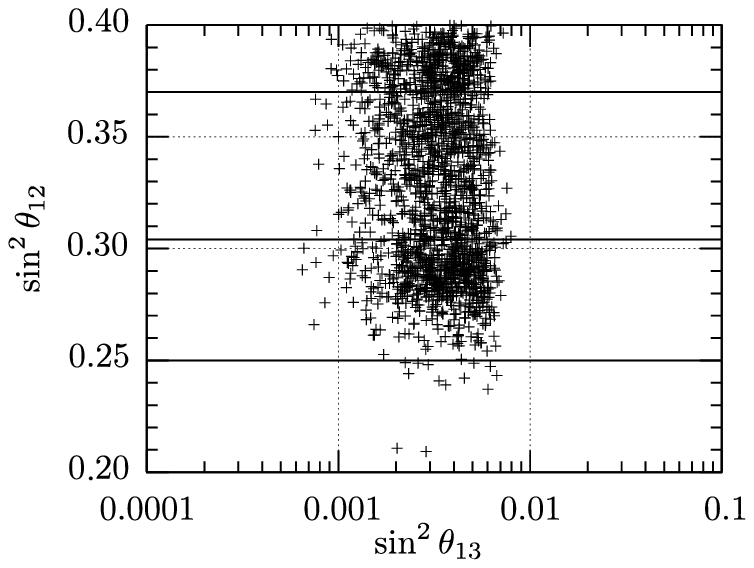}
\end{center}
\caption{Predicted PMNS mixing angles plotted in 
$\sin^2\theta_{13}$--$\sin^2\theta_{12}$ plane. Left and right plots are 
derived from cascade texture model I and II, respectively. Best-fit value 
with 3$\sigma$ interval 
of solar mixing angle $\sin^2\theta_{12}=0.304^{+0.066}_{-0.054}$ 
in~\cite{Schwetz:2008er} is also shown by horizontal lines. 
\bigskip}
\label{fig1}
\end{figure}
%%%%%%%%%%
In figure 1, we show the predicted PMNS mixing angles in 
$\sin^2\theta_{13}$--$\sin^2\theta_{12}$ plane. Left and right plots 
are derived from model I and II, respectively. Best-fit value with 
3$\sigma$ interval of solar mixing angle 
$\sin^2\theta_{12}=0.304^{+0.066}_{-0.054}$ in~\cite{Schwetz:2008er} 
is also shown by horizontal lines. For model I, $\sin^2\theta_{13}$ 
can take larger value than 0.01, which is also favored by recent 
neutrino oscillation data~\cite{Schwetz:2008er,Fogli:2008jx}, and also 
be much suppressed. On the other hand, for model II, predicted value 
of $\sin^2\theta_{13}$ is rather restricted. Since the contributions 
from charged lepton sector to the PMNS mixing angles have no 
difference between both models, the result implies that the other 
corrections which deviate from tri-bimaximal mixing in 
$\sin\theta_{13}$ are larger for model I, rather than model II. 
Predicted range of $\sin^2\theta_{12}$ has no significant difference 
between model I and II. In our case, the nearly tri-bimaximal 
generation mixing in neutrino sector dominates 12 mixing in PMNS 
matrix, and mixing from charged lepton sector give relatively small 
correction to $\sin^2\theta_{12}$. Thus the plots distributed around 
$\sin^2\theta_{12}=1/3$. It should be notified that the lower limit of 
$\sin^2\theta_{12}$ appear in the 3$\sigma$ interval. There are no 
particular correlation between predictions of $\sin^2\theta_{13}$ and 
$\sin^2\theta_{12}$. 
%%%%%%%%%% FIGURE.2
\begin{figure}[t]
\begin{center}
  \includegraphics[width=7cm,clip]{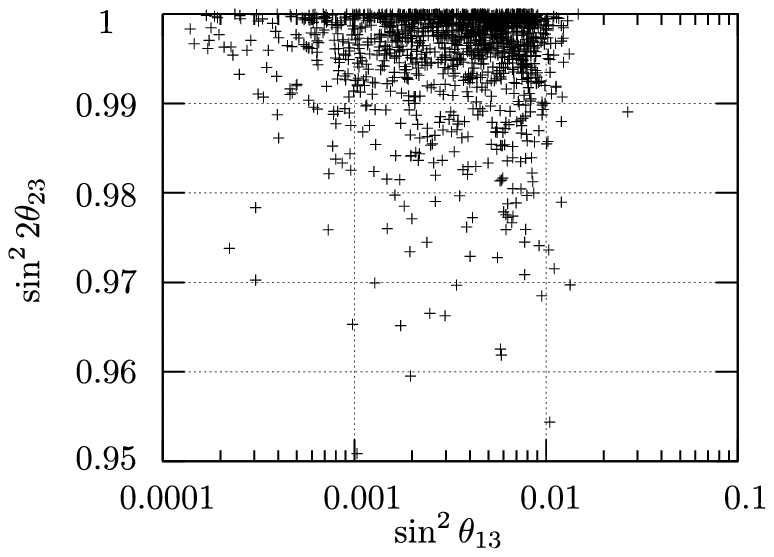}\hspace{1cm}
  \includegraphics[width=7cm,clip]{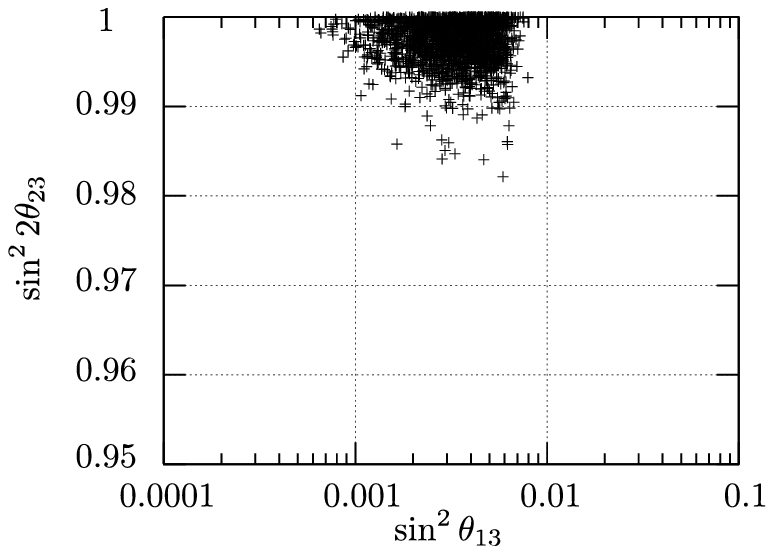}
\end{center}
\caption{Predicted PMNS mixing angles plotted in 
$\sin^2\theta_{13}$--$\sin^22\theta_{23}$ plane. Left and right plots are 
derived from cascade texture model I and II, respectively. 
\bigskip}
\label{fig2}
\end{figure}
%%%%%%%%%%
In figure 2, we also show the predicted PMNS mixing angles in 
$\sin^2\theta_{13}$--$\sin^22\theta_{23}$ plane. Left and right plots 
are derived from model I and II, respectively. In both models, 
prediction of $\sin^22\theta_{23}$ is mostly larger than 0.99, which 
is quite close to the best fit value. Due to the GJ factor in $M_e$, 
corrections to the atmospheric angle from charged lepton mixing is 
rather suppressed than the previous analysis of the cascade 
matrices~\cite{Haba:2008dp}. Finally, we present some figures showing 
correlations between a neutrino mass ratio, $m_1/m_2$, and each mixing angle 
in figure \ref{fig-new1}. Since the cascade model works only in the NH case, 
$m_2$ and $m_3$ are approximated by the solar and atmospheric scales. 
Therefore, the ratio $m_1/m_2$, strongly correlated to $m_1$, is chosen as 
the vertical axis. These analyses would be checked by the future neutrino 
experiments and cosmological bounds on its absolute mass. 
%%%%%%%%%% FIGURE.NEW
\begin{figure}%[thbp]
\begin{center}
  \includegraphics[width=6cm,clip]{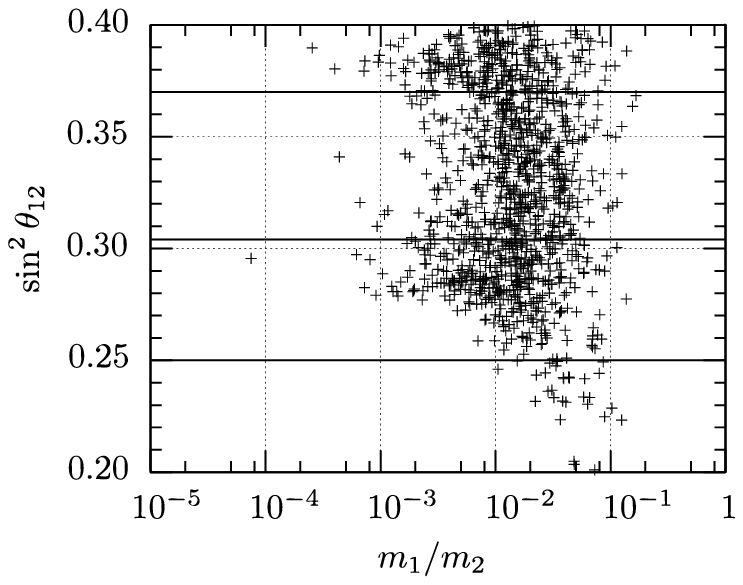}\hspace{1cm}
  \includegraphics[width=6cm,clip]{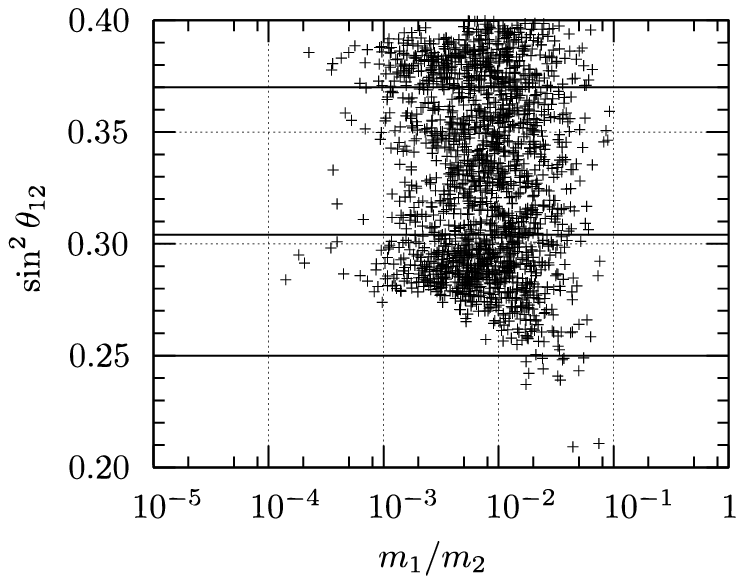}\\
  \includegraphics[width=6cm,clip]{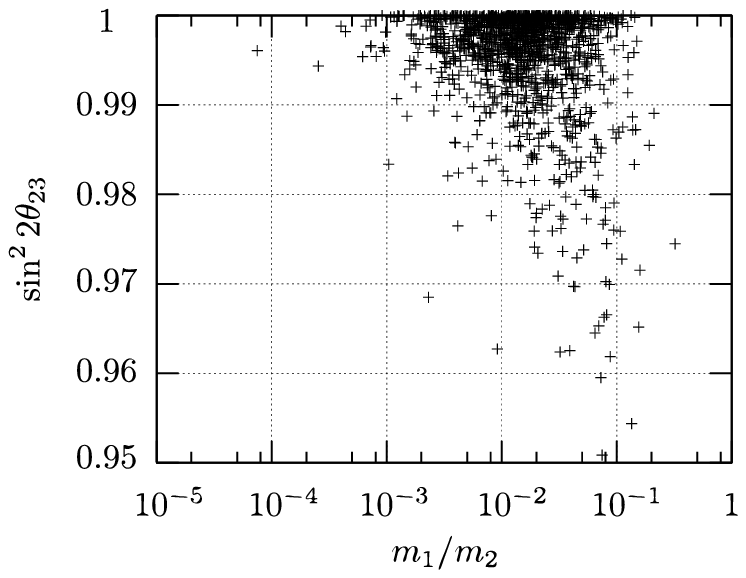}\hspace{1cm}
  \includegraphics[width=6cm,clip]{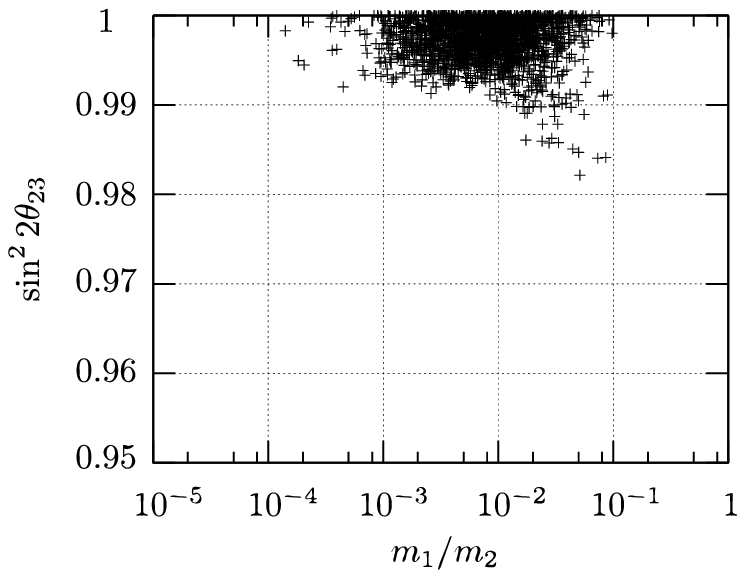}\\
  \includegraphics[width=6cm,clip]{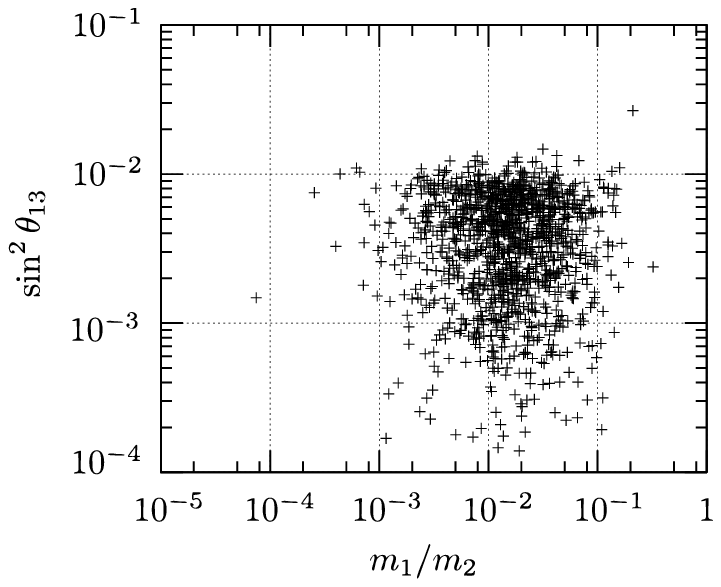}\hspace{1cm}
  \includegraphics[width=6cm,clip]{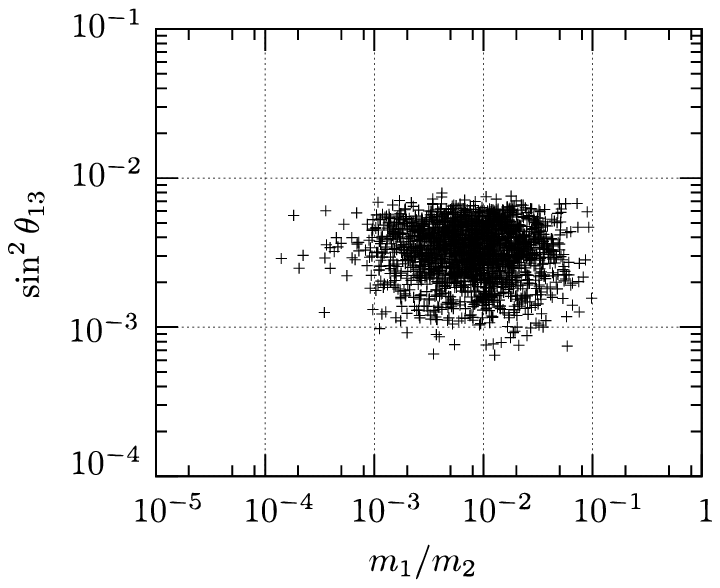}
\end{center}
\caption{Correlations between neutrino mass ratio and 
PMNS mixing angles in model I (left) and II (right). 
\bigskip}
\label{fig-new1}
\end{figure}
%%%%%%%%%%

%%%%%%%%%%%%%%%%%%%%%%%%%%%%%%%%%%%%
\subsection{Lepton flavor violation}
%%%%%%%%%%%%%%%%%%%%%%%%%%%%%%%%%%%%

Next, we estimate the branching ratios of flavor violating rare decays of 
charged leptons. Supersymmetric models generically induce sizable magnitudes of 
lepton flavor violation (LFV) because there exist additional sources of LFV, 
which are mass parameters of sleptons. Those flavor violating processes are 
radiatively generated depending on the structure of lepton mass matrices. We 
investigate the branching ratios of the rare decay processes, $l_i\rightarrow 
l_j\gamma$, in these cascade lepton mass matrices. 

For simplicity, we assume that soft SUSY breaking masses of sleptons are 
universal at the GUT scale, $\LamGUT$. Then the off-diagonal 
matrix elements are generated by radiative corrections from the Yukawa 
couplings of neutrinos~\cite{Borzumati:1986qx}. The one-loop renormalization 
group evolution induces the left-handed slepton masses, which are estimated as 
\begin{eqnarray}
  (m_l^2)_{ij}\sim\frac{1}{8\pi^2v^2}(3m_0^2+|a_0|^2)\sum_k(M_{\nu D}^\dagger)_{ik}
                (M_{\nu D})_{kj}\ln\left(\frac{|M_k|}{\LamGUT}\right)~~~
  (\mbox{for }i\neq j),
\end{eqnarray}
where $m_0$ and $a_0$ are the universal SUSY breaking mass and three-point 
coupling of scalar superpartners given at the GUT scale. The magnitude of these 
off-diagonal elements depends on the structure of neutrino Dirac mass matrix 
and the mass scale of right-handed Majorana neutrinos. 

\begin{figure}[]
\begin{center}
  \includegraphics[width=9cm,clip]{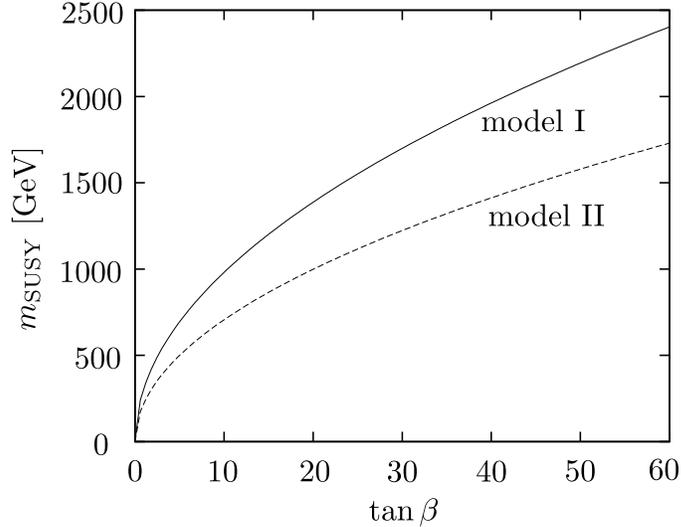}
\end{center}
\caption{Constraints for $\tan\beta$ and $m_{\rm SUSY}$ from experimental 
bounds of LFV processes. In the figure, solid (model I) and dashed (model II) 
lines show lower bounds of $m_{\rm SUSY}$ for particular values of $\tan\beta$. 
Above the line, all the LFV constraints~\eqref{lfvexp} are satisfied 
for each model. 
\bigskip}
\label{lfvfig}
\end{figure}

The branching ratio of $l_i\rightarrow l_j\gamma$ is roughly given by 
\begin{eqnarray}
  \mbox{Br}(l_i\rightarrow l_j\gamma)
  \simeq\frac{3\alpha}{2\pi}\frac{|(m_l^2)_{ij}|^2M_W^4}{\msusy^8}\tan^2\beta,
\end{eqnarray}
in the mass insertion approximation, where $\alpha$, $M_W$, and $\msusy$ are 
the fine structure constant, the $W$ boson mass, and a typical mass scale of 
superparticles circulating in one-loop diagrams, respectively. These branching 
ratios are calculated as 
\begin{eqnarray}
  \mbox{Br}(\mu\rightarrow e\gamma) 
   &\simeq& \frac{3\alpha}{8\pi^5}\lambda^{4d}B
            \left[\lambda^6\ln\left(\frac{|M_1|}{\LamGUT}\right)
                  +\lambda^4\ln\left(\frac{|M_2|}{\LamGUT}\right)
                  -\lambda^4\ln\left(\frac{|M_3|}{\LamGUT}\right)\right]^2, \\
  \mbox{Br}(\tau\rightarrow e\gamma) 
   &\simeq& \frac{3\alpha}{8\pi^5}\lambda^{4d}B
            \left[\lambda^6\ln\left(\frac{|M_1|}{\LamGUT}\right)
                  -\lambda^4\ln\left(\frac{|M_2|}{\LamGUT}\right)
                  +\lambda^3\ln\left(\frac{|M_3|}{\LamGUT}\right)\right]^2, \\
  \mbox{Br}(\tau\rightarrow\mu\gamma) 
   &\simeq& \frac{3\alpha}{8\pi^5}\lambda^{4d}B
            \left[\lambda^6\ln\left(\frac{|M_1|}{\LamGUT}\right)
                  -\lambda^2\ln\left(\frac{|M_2|}{\LamGUT}\right)
                  -\lambda\ln\left(\frac{|M_3|}{\LamGUT}\right)\right]^2,
\end{eqnarray}
for the model I, and 
\begin{eqnarray}
  \mbox{Br}(\mu\rightarrow e\gamma) 
   &\simeq& \frac{3\alpha}{8\pi^5}\lambda^{4d}B
            \left[\lambda^8\ln\left(\frac{|M_1|}{\LamGUT}\right)
                  +\lambda^5\ln\left(\frac{|M_2|}{\LamGUT}\right)
                  -\lambda^5\ln\left(\frac{|M_3|}{\LamGUT}\right)\right]^2, \\
  \mbox{Br}(\tau\rightarrow e\gamma) 
   &\simeq& \frac{3\alpha}{8\pi^5}\lambda^{4d}B
            \left[\lambda^8\ln\left(\frac{|M_1|}{\LamGUT}\right)
                  -\lambda^5\ln\left(\frac{|M_2|}{\LamGUT}\right)
                  +\lambda^4\ln\left(\frac{|M_3|}{\LamGUT}\right)\right]^2, \\
  \mbox{Br}(\tau\rightarrow\mu\gamma) 
   &\simeq& \frac{3\alpha}{8\pi^5}\lambda^{4d}B
            \left[\lambda^8\ln\left(\frac{|M_1|}{\LamGUT}\right)
                  -\lambda^2\ln\left(\frac{|M_2|}{\LamGUT}\right)
                  -\lambda\ln\left(\frac{|M_3|}{\LamGUT}\right)\right]^2,
\end{eqnarray}
for the model II, where $B\equiv(M_W/\msusy)^4\tan^2\beta$. 
Typical magnitudes of the branching ratios are shown in Tab.~\ref{tab11}. 
\begin{table}
\begin{center}
\begin{tabular}{|c|c|c|c|c|c|c|}
\hline
\multicolumn{7}{|c|}{Model I} \\
\hline
$d$  & $\frac{\mbox{Br}(\mu\rightarrow e\gamma)}{B}$ & $\frac{\mbox{Br}(\tau\rightarrow e\gamma)}{B}$ & $\frac{\mbox{Br}(\tau\rightarrow\mu\gamma)}{B}$ & $M_1$ [GeV] & $M_2$ [GeV] & $M_3$ [GeV]\\
\hline
\hline
$0$  & $2.66\times10^{-9}$ & $7.80\times10^{-10}$ & $3.99\times10^{-7}$ & $3.61\times10^{12}$ & $3.08\times10^{13}$ & $1.16\times10^{16}$\\
\hline
$1$ & $7.40\times10^{-12}$ & $7.60\times10^{-12}$ & $2.31\times10^{-9}$ & $1.80\times10^{10}$ & $1.59\times10^{12}$ & $5.98\times10^{14}$ \\
\hline
$2$ & $2.06\times10^{-14}$ & $1.28\times10^{-13}$ & $4.37\times10^{-11}$ & $4.22\times10^9$ & $8.19\times10^{10}$ & $3.08\times10^{13}$ \\
\hline
\end{tabular}\vspace{5mm}

\begin{tabular}{|c||c|c|c||c|c|c|}
\hline
\multicolumn{7}{|c|}{Model II} \\
\hline
$d$  & $\frac{\mbox{Br}(\mu\rightarrow e\gamma)}{B}$ & $\frac{\mbox{Br}(\tau\rightarrow e\gamma)}{B}$ & $\frac{\mbox{Br}(\tau\rightarrow\mu\gamma)}{B}$ & $M_1$ [GeV] & $M_2$ [GeV] & $M_3$ [GeV]\\
\hline
\hline
$0$  & $1.85\times10^{-10}$ & $3.57\times10^{-10}$ & $2.67\times10^{-6}$ & $3.61\times10^{11}$ & $3.08\times10^{13}$ & $5.11\times10^{16}$\\
\hline
$1$ & $4.95\times10^{-13}$ & $1.06\times10^{-15}$ & $1.66\times10^{-11}$ & $1.86\times10^9$ & $1.59\times10^{12}$ & $2.64\times10^{15}$ \\
\hline
$2$ & $1.33\times10^{-15}$ & $2.19\times10^{-15}$ & $1.54\times10^{-11}$ & $9.57\times10^8$ & $8.19\times10^{10}$ & $1.36\times10^{14}$\\
\hline
\end{tabular}
\end{center}
\caption{Typical magnitudes of branching ratios for lepton flavor violating rare decay process.}
\label{tab11}
\end{table}
In these analyses, 
$\LamGUT=2\times10^{16}$ GeV is 
taken.\footnote{Notice that $\tan\beta=38$ and $\msusy=500$ GeV have been 
taken in the numerical fit of quark and charged lepton masses with a high 
accuracy of those mass relations at GUT scale in the previous subsection 
but it was just for simplicity. The magnitude of the threshold corrections 
are important for those relations, and dependence of corrections on the 
overall SUSY scale, $\msusy$, is negligibly small as long as a model is 
discussed in a low scale SUSY breaking like in our case. In discussions 
of the LFV processes given here, we 
extend our consideration to be more general case, that is, we take $\tan\beta$ 
and $\msusy$ as free parameters of the models and estimate constraints on it 
from the LFV 
searches.} These results are compared with the current experimental 
upper bounds at $90\%$ confidence level~\cite{Brooks:1999pu,Hayasaka:2007vc}:
\begin{eqnarray}
  \mbox{Br}(\mu\rightarrow e\gamma)\leq1.2\times10^{-11},~~
  \mbox{Br}(\tau\rightarrow e\gamma)\leq1.2\times10^{-7},~~
  \mbox{Br}(\tau\rightarrow\mu\gamma)\leq4.5\times10^{-8}.
  \label{lfvexp}
\end{eqnarray}
The magnitude of the branching ratio for the $\tau\rightarrow e\gamma$ process 
predicted from the cascade model with $d=0$ is far below the experimental 
limit. Further, all values of the ratio with $d\geq1$ are also sufficiently 
smaller than the current bounds. On the other hand, the 
$\mu\rightarrow e\gamma$ and 
$\tau\rightarrow\mu\gamma$ decays are marginal to the present limit and would 
be observed in future LFV searches with relatively light superparticle 
spectrum. These experimental limit in turn constraints 
$\tan\beta$ and the mass scale of superparticles. 
The most severe constraint on the scale comes from the 
$\mu\rightarrow e\gamma$ for the model I with $d=0$ and 
$\tau\rightarrow\mu\gamma$ for the model II with $d=0$. They are 
$B\leq4.51\times10^{-3}$ and $B\leq1.68\times10^{-2}$ for the model I and II, 
respectively. 
They mean that the typical SUSY breaking scale has a lower bound as 
\begin{eqnarray}
  \msusy\geq1912\left(\frac{\tan\beta}{38}\right)^{1/2}\mbox{ GeV},
\end{eqnarray}
for the model I, and 
\begin{eqnarray}
  \msusy\geq1375\left(\frac{\tan\beta}{38}\right)^{1/2}\mbox{ GeV},
\end{eqnarray}
for the model II.\footnote{If we fix the value of $\msusy$ as $500$ GeV, the 
both magnitudes of the branching ratio for the $\mu\rightarrow e\gamma$ and 
$\tau\rightarrow\mu\gamma$ processes in models I and II with $d=0$ exceed the 
current experimental upper bound, and thus, this possibility is ruled out. 
Therefore, $d\geq1$ is required for our cascade textures. The parameter $d$ can 
 be related with the heaviest right-handed neutrino mass scale through 
$M_3=\lambda^{2(d+1)-x_2}v_u^2/\sqrt{|\Delta m_{31}^2|}$. Then, the 
constraints on the heaviest right-handed neutrino mass as 
$M_3\lesssim5.98\times10^{14}$ GeV for model I and 
$M_3\lesssim2.64\times10^{15}$ GeV for model II are obtained. It is expected 
that the lepton rare decay processes of $\mu\rightarrow e\gamma$ and 
$\tau\rightarrow\mu\gamma$ predicted from the minimal model I with $d=1$ would 
be observed in near future LFV searches. Those branching ratios become 
$\mbox{Br}(\mu\rightarrow e\gamma)=7.14\times10^{-12}$ and 
$\mbox{Br}(\tau\rightarrow\mu\gamma)=2.23\times10^{-9}$.} 
In figure~\ref{lfvfig}, the constraints are shown for model I and II. 
The solid (model I) and dashed (model II) lines show lower bounds 
of $m_{\rm SUSY}$ for particular values of $\tan\beta$. 
Above the line, all the LFV constraints~\eqref{lfvexp} are satisfied 
for each model. 
%%%%%%%%%% FIGURE.2
%%%%%%%%%%

%%%%%%%%%%%%%%%%%%%%%%%%%
\subsection{Leptogenesis}
%%%%%%%%%%%%%%%%%%%%%%%%% 

Next let us study CP violating phenomenology. Especially, we examine whether 
the thermal leptogenesis~\cite{Fukugita:1986hr} works in the cascade model. 
The CP asymmetry parameter in the right-handed neutrino, $R_i$, decay is given 
by 
\begin{eqnarray}
  \epsilon_i=\frac{\sum_j\Gamma(R_i\rightarrow L_jH)
    -\sum_j\Gamma(R_i\rightarrow L_j^cH^\dagger)}
  {\sum_j\Gamma(R_i\rightarrow L_jH)
    +\sum_j\Gamma(R_i\rightarrow L_j^cH^\dagger)},
\end{eqnarray}
where $L_i$ and $H$ denote the left-handed lepton and Higgs fields. An 
approximation for $\epsilon_i$ at low temperature is estimated as~\cite{LPG} 
\begin{eqnarray}
  \epsilon_1=\frac{1}{8\pi}\sum_{i\neq1}\frac{\mbox{Im}[(A_{i1})^2]}{|A_{11}|}F(r_i),
\end{eqnarray}
where $r_i\equiv|M_i/M_1|^2$, 
$A\equiv(DM_{\nu D}M_{\nu D}^\dagger D^\dagger)/v_u^2$, and $D$ is the diagonal 
phase matrix to make the eigenvalues $M_i$ real and positive. 
The formula is given in the diagonal basis of right-handed Majorana mass matrix 
with real positive eigenvalues and are reasonably accurate even at higher 
temperatures. The function $F$ denotes contributions from the one-loop vertex 
and self-energy corrections, 
\begin{eqnarray}
  F(x)=\sqrt{x}\left[\frac{2}{1-x}-\ln\left(1+\frac{1}{x}\right)\right], 
\end{eqnarray}
which is well approximated by $-3/\sqrt{x}$ for $x\gg 1$. 
The relevant quantities for the CP asymmetry parameter are given in 
Tab.~\ref{tab12}.
\begin{table}
  \begin{center}
    \begin{tabular}{|c|c|c|c|c|}
      \hline
      \multicolumn{5}{|c|}{Model I} \\
      \hline
      \hline
      $|A_{11}|$ & $|A_{21}|$ & $|A_{31}|$ & $|M_1/M_3|$ & $|M_1/M_2|$ \\
      \hline
      \hline
      3$\lambda^6$ & $\lambda^6$ & $\lambda^3$ & $\lambda^7$ & $\lambda^3$ \\
      \hline
    \end{tabular}\hspace{2mm}
    \begin{tabular}{|c|c|c|c|c|}
      \hline
      \multicolumn{5}{|c|}{Model II} \\
      \hline
      \hline
      $|A_{11}|$ & $|A_{21}|$ & $|A_{31}|$ & $|M_1/M_3|$ & $|M_1/M_2|$ \\
      \hline
      \hline
      3$\lambda^8$ & $\lambda^8$ & $\lambda^4$ & $\lambda^9$ & $\lambda^4$ \\
      \hline
    \end{tabular}
  \end{center}
  \caption{The relevant quantities for the CP asymmetry parameter.}
  \label{tab12}
\end{table}
By utilizing these quantities, the CP asymmetry parameter for each model can be 
calculated as 
\begin{eqnarray}
  \epsilon_1\simeq
  \left\{
    \begin{array}{ll}
      \frac{-1}{8\pi}
      \left[\lambda^9\sin(\theta_2-\theta_1)
        +\lambda^7\sin(\theta_3-\theta_1)
      \right] & \mbox{ for model I,}\\[2mm]
      \frac{-1}{8\pi}
      \left[\lambda^{12}\sin(\theta_2-\theta_1)
        +\lambda^9\sin(\theta_3-\theta_1)
      \right] & \mbox{ for model II.}
    \end{array}
  \right.\label{CPasym}
\end{eqnarray}
where $\theta_i=\arg(M_i)$. It is found that the second term is dominant unless 
the relevant argument is taken as a specific value such as zero. We here define 
the resultant CP asymmetry, $\etaCP$, as the ratio of the lepton asymmetry to 
the photon number density, $n_\gamma$, 
\begin{eqnarray}
  \etaCP =\frac{135\zeta(3)}{4\pi^4}\frac{\kappa s}{g_\ast}
  \frac{\epsilon_1}{n_\gamma},
\end{eqnarray}
where $\kappa$, $s$, and $g_\ast$ are the efficiency factor, entropy density, 
and the effective number of degrees of freedom in thermal equilibrium. They are 
given by~\cite{LPG2} 
\begin{eqnarray}
  \kappa^{-1} &\simeq& \frac{3.3\times10^{-3}\mbox{ eV}}{\meff}
  +\left(\frac{\meff}{5.5\times10^{-4}\mbox{ eV}}
  \right)^{1.16}, \\
  s          &=&      7.04n_\gamma, \\
  g_\ast      &=&      228.75.
\end{eqnarray}
The $\meff$ in the efficiency factor is the effective light neutrino mass 
defined as $\meff\equiv|(M_{\nu D}^\dagger M_{\nu D})_{11}/M_1|$. It is known 
that the efficiency depends only on $\meff$ when $|M_1|\ll10^{14}$ GeV, which 
is realized in both models. Finally, the baryon asymmetry of the universe, 
$\eta_B$, is transferred via spharelon interactions as $\eta_B=-8\etaCP/23$. 
As the result, the baryon asymmetry in our model is predicted as 
\begin{eqnarray}
  \eta_B\sim
  \left\{
    \begin{array}{ll}
      4.7\times10^{-11}\sin\theta_B & \mbox{ for model I,}\\
      1.3\times10^{-11}\sin\theta_B & \mbox{ for model II,}
    \end{array}
  \right.
\end{eqnarray}
where $\theta_B\equiv\theta_3-\theta_1$ and we take account only 
of the leading term in \eqref{CPasym}. These results are compared with the 
current observational data at $68\%$ confidence level from the WMAP 7-years 
mean result in the standard $\Lambda$CDM model, 
$\eta_B=(6.19\pm0.15)\times10^{-10}$~\cite{Komatsu:2010fb}. 
Although the prediction seems to be a little small, note that the above our 
naive estimation of $\eta_B$ does not involve the effects from combination 
of ${\cal O}(1)$ coefficients which generally exist in 
$M_{\nu D}$ and $M_R$. When the combination of ${\cal O}(1)$ coefficients 
generates ${\cal O}(10)$ (${\cal O}(50)$) enhancement factor of the asymmetry 
for model I (II), then $\eta_B$ can be consistent with the experimental bound 
as long as the relevant CP violation $\theta_B$ is maximally large. 
It is interesting that our models, which is constrained by only the neutrino 
mass spectra and PMNS structure, can lead to favored magnitude of the 
baryon asymmetry via leptogenesis.

%%%%%%%%%%%%%%%%%%%%%
\section{Discussions}
%%%%%%%%%%%%%%%%%%%%%

At the end of this paper, we discuss about a realization of our cascade model 
in SUSY $SU(5)$ GUT. Especially, we give some comments on flavor symmetry 
behind the model and extensions of SUSY $SU(5)$ GUT. 

%%%%%%%%%%%%%%%%%%%%%%%%%%%%%
\subsection{Flavour symmetry}
%%%%%%%%%%%%%%%%%%%%%%%%%%%%%

We have presented the texture analyses and shown some phenomenological results. 
These textures are described by the cascade form for the neutrino Dirac mass 
(Yukawa) matrix and the H.C. one for the charged lepton and down quark one. 
The effects from the right-handed Majorana and up quark mass matrices to the 
PMNS and CKM structures should be enough small not to spoil the 
experimentally favored mixing angles induced from the cascade form of neutrino 
Dirac and H.C. form of down quark mass matrices. The cascade (H.C.) contains 
two (three) step hierarchies. A smaller factor is concerned with the 1st 
generation and the other with 2nd one. Moreover, the current neutrino 
experimental data would suggest that the coefficients of effective mass 
operators are correlated to each other. These non-trivial features imply some 
implements introduced in fundamental theory beyond the SM. 

An introduction of flavor symmetries is one of such implements. Actually, a 
simple realization of cascade hierarchy has been shown in~\cite{Haba:2008dp} 
based on an $U(1)$ flavor symmetry. In the realization, three gauge 
singlet scalars, $\phi_i$, are introduced in addition to the fundamental SM 
fields. These field including the SM field are charged under the $U(1)$ flavor 
symmetry. A simple quantum number assignment of $U(1)$ flavor symmetry and a 
requirement of the same magnitude of expectation values, $\langle\phi_1\rangle 
\simeq\langle\phi_2\rangle\simeq\langle\phi_3\rangle\equiv\lambda\Lambda$, lead
 to cascade texture as~\eqref{cas}, where $\delta=\lambda^{m+1}$ and $m$ is an 
arbitrary positive integer. A significant feature of the flavor model is 
$\delta\leq \lambda^2$ in \eqref{cas}, which is suitable for the neutrino Dirac
 mass matrix as discussed in section~\ref{sec-Neutrino-sector}. In addition to 
the cascade form of neutrino Dirac mass matrix, a realization of H.C. form of 
charged lepton and down-type quark mass matrices could be obtained by extending
 the gauge singlet Higgs sector and number of $U(1)$ flavor symmetry. It is 
worth proceeding a study about simple realization of our model in terms of 
abelian flavor symmetry.\footnote{Some complicated mechanisms, such as discrete
 flavor symmetries, could be needed for the alignment of cascade and H.C. forms
. Although a complete flavor model would include additional flavor mechanism, 
we focus only on the realization of hierarchical structure of the mass matrices
 in this analysis.} 

For the analysis, let us introduce $U(1)_F\times U(1)_{F'}\times Z_3$ 
flavor symmetry and additional fields $\phi_{f}$, $\phi'_{f}$, $\phi''_{f}$, 
$\phi_{f'}$, $\phi'_{f'}$, $\phi''_{f'}$ and $\phi_z$, which are neutral under 
the $SU(5)$ gauge symmetry. SM fermions and Higgs fields are involved in 
$SU(5)$ representations; here the matter fields are denoted by $10_i$, 
$\bar 5_i$ and $1_i$ ($i=1,2,3$). We assume that the $U(1)_F\times U(1)_{F'}$ 
flavor symmetry is broken by the Higgs vacuum extension values 
$\langle \phi_{f}\rangle\simeq \langle \phi'_{f}\rangle \simeq 
 \langle \phi''_{f}\rangle \simeq \langle  \phi_{f'}\rangle 
\simeq \langle \phi'_{f'}\rangle \simeq \langle \phi''_{f'}\rangle \simeq 
\lambda \Lambda$, where $\Lambda$ is the cutoff scale of the theory. 
Also the discrete $Z_3$ is broken by 
$\langle \phi_z\rangle \simeq M'<\lambda\Lambda$. 
Although the vacuum expectation values would be determined by possible dynamics 
of the Higgs sector, those are simply adopted in the analysis. To give suitable 
$U(1)_F\times U(1)_{F'}\times Z_3$ charge assignments for the $\phi$'s and 
matter fields\footnote{We take up- and down-type Higgs fields are neutral under 
$U(1)_F\times U(1)_{F'}$, and have $1/3$ charges of $Z_3$ ($mod$ 1).}, one can 
lead to effective mass matrices of the quark and lepton fields via 
higher-dimensional operators suppressed by $\Lambda$. 

\begin{table}[t]
  \centering
  \begin{tabular}{c||ccc|ccc|ccc|ccc|ccc}
%\hline \hline
    &$10_1$&$10_2$&$10_3$&$\bar 5_1$&$\bar 5_2$&$\bar 5_3$&$1_1$&$1_2$&$1_3$
    &$\phi_f$&$\phi_f'$&$\phi_f''$&$\phi_{f'}$&$\phi_{f'}'$&$\phi_{f'}''$
    \\ \hline \hline 
    $U(1)_F$&6&4&0&5&2&-3&9&4&0&1&-2&-4&0&0&0\\
    $U(1)_{F'}$&-5&3&0&-6&-4&0&5&5&2&0&0&0&-2&3&-4\\
%\hline \hline
  \end{tabular}
\caption{An example of phenomenologically viable charge assignment of 
$U(1)_{F}\times U(1)_{F'}$ flavor symmetry. }
\label{tab-u1}
\end{table}
In Tab.~\ref{tab-u1}, an example of phenomenologically viable 
$U(1)_F\times U(1)_{F'}$ charge assignment is shown. For $Z_3$ charges, 
$\phi_{f,f'}$ are neutral and $10_i$, $\bar 5_i$ and $\phi_z$ have 
$1/3$ ($mod$ 1). In this case, one can obtain the following hierarchical 
structure in the effective mass matrices: 
\begin{eqnarray}
  M_u&\simeq &
  \left(
    \begin{array}{ccc}
      \lambda^{8} & \lambda^6 & \lambda^6 \\
      \lambda^6       & \lambda^4 & \lambda^4 \\
      \lambda^6       & \lambda^4 & 1
    \end{array}
  \right)v_u, ~~
  M_d\;\simeq \;
  \left(
    \begin{array}{ccc}
      \lambda^{8} & \lambda^3 & \lambda^3 \\
      \lambda^3       & \lambda^2 & \lambda^2 \\
      \lambda^3       & \lambda^2 & 1
    \end{array}
  \right)\lambda^2 v_d,\\
  M_{\nu D}&\simeq &
  \left(
    \begin{array}{ccc}
      \lambda^3 & \lambda^3 & \lambda^3 \\
      \lambda^3       & \lambda & \lambda \\
      \lambda^3       & \lambda & 1
    \end{array}
  \right)\lambda^3 v_u, ~~
  M_R\;\simeq \;
  \left(
    \begin{array}{ccc}
      \lambda^7 & \lambda^7 & \lambda^7 \\
      \lambda^7       & \lambda^4 & \lambda^4 \\
      \lambda^7       & \lambda^4 & 1
    \end{array}
  \right)\lambda M'. 
\end{eqnarray}
For the quark (and the charged lepton) sector, H.C. mass matrices are 
obtained. For the neutrino sector, Dirac mass matrix and mass 
eigenvalues of $M_R$ correspond to model I in section~\ref{sec-genmix}. 
Atmospheric neutrino mass scale determines the magnitude of Majorana mass 
scale as $M'\simeq 10^{13}$~GeV. Compared to model I, mixing between 1st and 
2nd generation of $M_R$ is larger in this case. We numerically checked 
that prediction of the PMNS angles in this case can satisfy the experimental 
constraints as the case of model I. The analysis implies that the cascade 
types of hierarchical structure in quark and 
lepton mass matrices can be obtained by simple flavor mechanisms. 

%%%%%%%%%%%%%%%%%%%%%%%%%%%%%%%%%%%%%%%%%%
\subsection{Extension of SUSY $SU(5)$ GUT}
%%%%%%%%%%%%%%%%%%%%%%%%%%%%%%%%%%%%%%%%%%

Cascade and H.C. textures are suitable for comprehensive description of 
hierarchical structure in quark and lepton mass matrices. This implies that 
the matrices are compatible with GUT, where quark and lepton flavor structure 
is generally related. In $SU(5)$ GUT, unification of down-type quark and 
charged lepton leads to constraints between elements of $M_d$ and $M_e$. 
In our texture analysis, the GJ factor~\cite{Georgi:1979df} is minimally 
introduced as in~\eqref{GJ}. Let us give complement discussion about 
the modification of the relation. 

A significant feature of the cascade hierarchy in $SU(5)$ GUT is that 
the CKM and PMNS matrices should be mainly controlled by mixing 
structure of down-type quark and neutrino sectors, respectively. 
In other words, mixing structure of $M_d$ is close to the CKM matrix, 
and neutrino sector leads to tri-bimaximal generation mixing at the 
leading order. Details of the relation between $M_d$ and $M_e$ 
controls mixing structure of $M_e$, which gives correction to 
PMNS mixing angles as studied in section~\ref{sec-clepton}. 

Relation between $M_d$ and $M_e$ is determined by details of high-energy 
models, and variety of GUT models could be considered. GUT scale fermion masses 
extrapolated from low-energy experimental data give implications of the 
relation of the mass matrices and GUT models. As mentioned, threshold effects 
which depend on the superparticle spectrum could play an important role to 
determine the GUT scale fermion masses (Yukawa couplings) 
in supersymmetric scenario~\cite{deltab,gutph}. In the recent 
analysis~\cite{Antusch:2008tf}, for example, the possible relation 
between down-type quark and charged lepton masses (and corresponding 
GUT model) has been studied with several SUSY breaking scenarios. 

Cascade hierarchies is naturally compatible with the several possibilities. 
Modification of the GJ relation $(m_e/m_d,m_\mu/m_s,m_\tau/m_b)\simeq (1/3,3,1)$ 
leads to different mixing structure of $M_e$ from our study, that is, thus 
prediction of PMNS angles would be somewhat changed depends on details of the 
mass relation. For example, mass relation 
$(m_e/m_d,m_\mu/m_s,m_\tau/m_b)\simeq (3/8,6,3/2)$ 
is compatible with typical SUSY breaking scenarios~\cite{Antusch:2008tf}. 
With the ordinary matter assignment of $SU(5)$ representation, 
H.C. structure of $M_d$ and $M_e$ could be minimally modified by CG 
coefficients through dimension--five interaction involving Higgs fields, 
as follows: 
\begin{eqnarray}
  M_d\simeq   \left(
    \begin{array}{ccc}
      \epsilon_d & \delta_d & \delta_d  \\
      \delta_d       & \lambda_d & \lambda_d \\
      \delta_d       & \lambda_d & 1
    \end{array}
  \right)\xi_d v_d,~~
  M_e\simeq   \left(
    \begin{array}{ccc}
      \epsilon_d & 3/2\delta_d & \delta_d  \\
      3/2\delta_d       & 6\lambda_d & \lambda_d \\
      \delta_d       & \lambda_d & 3/2
    \end{array}
  \right)\xi_d v_d.
\end{eqnarray}
In this case, mixing structure of charged lepton mass matrix is 
changed from our analysis: at the leading order, mixing angles between 
1st--2nd, 2nd--3rd and 1st--3rd generations are multiplied 3/4, 2/3, and 3/2 
by the minimal GJ case~\eqref{GJ}, respectively. As a result, prediction of 
PMNS mixing angles are slightly modified through mixing structure of $M_e$. 

Future progress of experimental searches is expected to give precise 
information about low-energy flavor structure and also SUSY parameters. 
Combined analyses of the fermion flavor structure and SUSY parameters would 
give key ingredients to reveal high-energy flavor origin. Examination of 
several types of cascade hierarchies and comparison between GUT models are 
left to our further study.

%%%%%%%%%%%%%%%%%
\section{Summary}
%%%%%%%%%%%%%%%%%

We have presented texture analyses based on cascade form. The neutrino 
Dirac mass matrix of a cascade form can lead to the tri-bimaximal 
generation mixing at the leading order in the framework of seesaw 
mechanism. On the other hand, the down quark mass matrix of a hybrid 
cascade form can reproduce the CKM structure. These facts give us a 
motivation to study cascade textures in a grand unified theory. 

We have embedded such experimentally favored mass textures into a SUSY 
$SU(5)$ GUT, which gives a relation between the down quark and charged 
lepton mass matrices. This relation constrains the structure of charged 
lepton mass matrix to a hybrid cascade form. We have taken the right-handed 
Majorana mass matrix as a form which gives enough small corrections to the PMNS 
structure not to spoil generation mixing structures induced from the neutrino 
Dirac mass matrix of the cascade form. The mass matrix of up-type quarks is 
also supposed to be a hybrid cascade form in our analyses. Related 
phenomenologies, such as lepton flavor violating processes and leptogenesis, 
have been also investigated in addition to lepton mixing angles in two typical 
types of model. 

For the lepton mixing angles, the both models described by cascade and hybrid 
cascade textures give an upper bound on the $\sin^2\theta_{13}$. The value of 
$\sin^2\theta_{13}$ is determined by summation of collections from the charged 
lepton and right-handed neutrino sectors, and properties of cascade texture of 
neutrino Dirac mass matrix. The value of upper bound in model I, which is 
$\sin^2\theta_{13}\lesssim0.01$, is larger than one in model II. It might be 
checked in upcoming neutrino oscillation experiments. Predicted range of 
$\sin^2\theta_{12}$ has no significant difference between model I and II. Since 
correction from charged lepton sector are relatively small, the generation 
mixing in the neutrino sector dominates $\theta_{12}$. The predictions of 
$\sin^22\theta_{23}$ are mostly larger than 0.99 in both models. Finally, some 
correlations between a neutrino mass ratio and each mixing angle have been also 
presented. 

In estimations of lepton flavor violating rare decay processes, it has been 
shown that the most severe constraint on a typical SUSY scale correlating with 
$\tan\beta$ comes from $\mu\rightarrow e\gamma$ process for the model I and 
$\tau\rightarrow\mu\gamma$ for the model II. We have also examined whether the 
thermal leptogenesis works in both cascade models. Enough baryon asymmetry via 
the thermal leptogenesis cannot be generated because of a relatively large 
hierarchy among the right-handed Majorana masses. Therefore, we need other 
source of baryon asymmetry in order to reproduce the observed values. 

%%%%%%%%%%%%%%%%%%%%%%%%%%%%%
\subsection*{Acknowledgement}
%%%%%%%%%%%%%%%%%%%%%%%%%%%%

The work of R.T. is supported by the DFG-SFB TR 27.

\appendix

%%%%%%%%%%%%%%%%%%%%%%%%%%%%%%%%%%%%%%%%%%%%%%%%%%%%%%%%
\section{Constraints on structure of non-diagonal $M_R$}
%%%%%%%%%%%%%%%%%%%%%%%%%%%%%%%%%%%%%%%%%%%%%%%%%%%%%%%%

We discuss constraints on the structure of non-diagonal $M_R$ in this section. 
In the section 3.4.2, we have considered a non-diagonal $M_R$ and effects from 
off-diagonal elements. 

We have defined the diagonalized mass matrix of the right-handed neutrino, 
$D_R$, as in~\eqref{DR} and an unitary matrix, $U_{\nu R}$, which diagonalize 
the $M_R$, as in~\eqref{UnuR}. The resultant neutrino mass after the seesaw 
mechanism and operating the tri-bimaximal mixing is given 
in~\eqref{off-neu-mm}. Each matrix element is written down as
 \begin{eqnarray}
  \mathcal{M}_{11}
   &\simeq& \frac{\lambda^{2d}v_u^2}{6M}
            [1+4\lambda^{2d_1-x_2}+4\lambda^{d_1}\theta_{R,23}
             +\lambda^{-x_2}\theta_{R,23}^2 \nonumber \\
   &      & \phantom{\frac{\lambda^{2d}v_u^2}{6M}[}
            +\lambda^{-x_1}(4\lambda^{2d_1}\theta_{R,12}
            -4\lambda^{d_1}\theta_{R,12}\theta_{R,13}+\theta_{R,13}^2)], \\
  \mathcal{M}_{22} 
   &\simeq& \frac{\lambda^{2d}v_u^2}{M}
            [3\lambda^{2d_1-x_1}+\frac{1}{3}+\frac{\lambda^{2d_1-x_2}}{3}
             +\lambda^{-x_2}(-\frac{2\lambda^{d_1}}{3}\theta_{R,23}+\theta_{R,23}^2) 
            \nonumber \\
   &      & \phantom{\frac{\lambda^{2d}v_u^2}{M}[}
            +\lambda^{-x_1}(-2\lambda^{d_1}\theta_{R,13}+\theta_{R,13}^2)],\\
  \mathcal{M}_{33}
   &\simeq& \frac{\lambda^{2d}v_u^2}{M}
            [2\lambda^{2d_2-x_2}+\frac{1}{2}
             +\lambda^{-x_2}(2\lambda^{d_2}\theta_{R,23})+\frac{\theta_{R,23}^2}{2}
                           -4\lambda^{d_2}\theta_{R,12}\theta_{R,13} \nonumber \\
   &      & \phantom{\frac{\lambda^{2d}v_u^2}{M}[}
            +\lambda^{-x_1}(2\lambda^{2d_2}\theta_{R,12}^2
                          +\frac{1}{2}\theta_{R,13}^2)], \\
  \mathcal{M}_{12}
   &\simeq& -\frac{\lambda^{2d}v_u^2}{3\sqrt{2}M}
            [1+\lambda^{d_1-x_2}\theta_{R,23}
             +3\lambda^{-x_1}(-2\lambda^{d_1}\theta_{R,12}
                            +\lambda^{d_1}\theta_{R,13})], \\
  \mathcal{M}_{23}
   &\simeq& \frac{\lambda^{2d}v_u^2}{\sqrt{6}M}
            [1-2\lambda^{d_1+d_2-x_2}+2\lambda^{d_2-x_2}\theta_{R,23} \nonumber \\
   &      & \phantom{\frac{\lambda^{2d}v_u^2}{\sqrt{6}M}[}
            +3\lambda^{-x_1}(2\lambda^{d_1+d_2}\theta_{R,12}-\lambda^{d_1}\theta_{R,13}
                           -2\lambda^{d_2}\theta_{R,13}\theta_{R,12}
                           +\theta_{R,13}^2)], \\
  \mathcal{M}_{13}
   &\simeq& -\frac{\lambda^{2d}v_u^2}{2\sqrt{3}M}
            [1+\lambda^{-x_2}(4\lambda^{d_1+d_2}+2\lambda^{d_2}\theta_{R,23}
                            +\theta_{R,23}^2) \nonumber \\
   &      & \phantom{-\frac{\lambda^{2d}v_u^2}{2\sqrt{3}M}[}
            -\lambda^{-x_1}(2\lambda^{d_2}\theta_{R,12}\theta_{R,13}
                          -4\lambda^{d_1+d_2}\theta_{R,12}^2-\theta_{R,13}^2)],
 \end{eqnarray}
where we have omitted terms which are trivially small compared with other 
terms. We require that the magnitudes of leading order of each term in this 
mass matrix are the same one as in the case of diagonal $M_R$ case, because the 
tri-bimaximal mixing can be realized at the leading order. This 
requirement leads to constraints on the mixing angles, $\theta_{R,ij}$, as 
 \begin{eqnarray}
  \theta_{R,13} &<& \frac{3}{2}\lambda^{d_1},~\frac{1}{3}\lambda^{-d_1+x_1},~
                  2\lambda^{d_2+(x_1-x_2)/2},~\frac{1}{\sqrt{3}}\lambda^{x_1/2}, \\
  \theta_{R,23} &<& \lambda^{-d_1+x_2},~\lambda^{d_2},~
                  \frac{1}{2}\lambda^{-d_2+x_2},~\lambda^{x_2/2}, \\
  \theta_{R,12} &<& \frac{1}{6}\lambda^{-d_1-d_2+x_1},~\lambda^{(x_1-x_2)/2}, \\
  \theta_{R,12}\theta_{R,13} &<& \lambda^{d_2+x_1-x_2},~\frac{1}{6}\lambda^{-d_2+x_1}.
 \end{eqnarray}
If we fix the values of $(d_1,d_2,x_1,x_2)$, which must be satisfied the 
conditions \eqref{constraint1}, we can determine the structure leading to 
maximal collections to the PMNS structure and neutrino mass spectra as shown in 
Tabs.~\ref{tab9} and~\ref{tab10}.

%%%%%%%%%%%%%%%%%%%%%%%%%%%

\end{document}